\documentclass{aastex}	        

\usepackage{emulateapj5}	        

\newcommand{\hi}{\hbox{H~I}}
\newcommand{\hii}{\hbox{H~II}}
\newcommand{\lsun}{\hbox{$L_{\odot,B}$}}
\newcommand{\msun}{\hbox{$M_{\odot}$}}

\newcommand{\otwo}{\hbox{[O II]$\lambda 3727$}}

\newcommand{\hgamma}{\hbox{H$\gamma$}}
\newcommand{\othreea}{\hbox{[O III]$\lambda 4363$}}
\newcommand{\hbeta}{\hbox{H$\beta$}}
\newcommand{\othree}{\hbox{[O III]$\lambda\lambda 4959,5007$}}

\newcommand{\othreec}{\hbox{[O III]$\lambda 5007$}}
\newcommand{\halpha}{\hbox{H$\alpha$}}

\newcommand{\ntwo}{\hbox{[N II]$\lambda\lambda 6548,6583$}}
\newcommand{\stwo}{\hbox{[S II]$\lambda\lambda 6716,6731$}}
\newcommand{\stwoa}{\hbox{[S II]$\lambda 6716$}}
\newcommand{\stwob}{\hbox{[S II]$\lambda 6731$}}

\journalinfo{Astron. Journal, scheduled Jan. 2003}	

\shortauthors{Lee et al.}
\shorttitle{Clues to Galaxy Evolution. I.}
\slugcomment{v2: includes revisions after proofs; emulateapj mode}

\begin{document}

\title{
Uncovering Additional Clues to Galaxy Evolution. \\
I.  Dwarf Irregulars in the Field
}

\author{
Henry Lee$\,$\altaffilmark{1,2}, 
Marshall L. McCall$\,$\altaffilmark{2,3,4},
Robin L. Kingsburgh$\,$\altaffilmark{2,4}, \\
Robert Ross$\,$\altaffilmark{2}, and
Chris C. Stevenson$\,$\altaffilmark{2,3,5}
}

\altaffiltext{1}{
Max-Planck-Institut f\"ur Astronomie, 
K\"onigstuhl 17, D--69117, Heidelberg, Germany.
E-mail: {\tt lee@mpia.de}  
}

\altaffiltext{2}{Department of Physics \& Astronomy, York University,
4700 Keele St., Toronto, ON, Canada  M3J 1P3.
}

\altaffiltext{3}{Visiting Astronomer, Steward Observatory, 
operated by The University of Arizona.
}

\altaffiltext{4}{Visiting Astronomer, Observatorio Astron\'omico 
Nacional (San Pedro M\'artir), operated by the Instituto de 
Astronom\'{\i}a, Universidad Nacional Auton\'oma de M\'exico.
} 

\altaffiltext{5}{Department of Physics \& Physical
Oceanography, Memorial University of Newfoundland, St. John's,
Newfoundland, Canada  A1B 3X7.
} 

\begin{abstract}				
In order to recognize environmental effects on the evolution of 
dwarf galaxies in clusters of galaxies, it is first necessary to
quantify the properties of objects which have evolved in relative 
isolation.
With oxygen abundance as the gauge of metallicity, two key diagnostics
of the evolution of dwarf irregular galaxies in the field are
re--examined: 
the metallicity--luminosity relationship and the 
metallicity--gas fraction relationship.
Gas fractions are evaluated from the masses of luminous components
only, i.e., constituents of the nucleogenetic pool.
Results from new optical spectroscopy obtained for \hii\ regions in 
five dwarf irregular galaxies in the Local Volume are incorporated
into a new analysis of field dwarfs with [O~III]$\lambda$4363
detections and good distances.
The updated fit to the metallicity--luminosity relationship is 
consistent with results reported in the literature.
The fit to the metallicity--gas fraction relation shows an excellent 
correlation consistent with expectations of the simple ``closed box''
model of chemical evolution. 
The simplest explanation consistent with the data is that flow
rates are zero, although the observations allow for the
possibility of modest flows. 
The derived oxygen yield is one-quarter of the value for the solar
neighbourhood. 
\end{abstract}


\section{Introduction}
\label{sec_intro}

Dwarf galaxies make the largest contribution to galaxies by
number and type
(e.g., \citealp{vcc,ferguson89,mdc97,secker97,mateo98,phillipps98,dc00})
and are considered to be the building blocks in hierarchical
scenarios of galaxy formation.
Being metal--poor, they represent the systems which have evolved least
since the Big Bang.
Although some dwarf galaxies are exceedingly faint, they may contain
large amounts of dark matter in proportion to their luminous masses.

Dwarf irregular galaxies (dIs), which are the focus of this paper,
are characterized by an irregular visual appearance, 
large gas reservoirs (e.g., \citealp{rh94,hoffman96}),
low dust content (e.g., see review by \citealp{calzetti01}), 
a lack of distinct spiral arms, and
star formation scattered throughout the body of the galaxies
(see also reviews by \citealp{gh84,hg86,hg89,hunter97}).
\hii\ regions provide a way to judge the total metallicity, 
because measures of oxygen abundance within these \hii\ regions 
probe the most recent period of chemical enrichment 
(e.g., \citealp{dinerstein90,shields90}).
Oxygen abundances within dIs do not vary greatly with galactocentric
radius (e.g., \citealp{ks96,devost97,ks97}).
Blue compact dwarf galaxies (BCDs) contain very bright central \hii\
regions. 
Here, dIs are chosen for study over BCDs, because the contribution of
light from young stars in dIs is much less than the light from the
underlying old stellar population (e.g., \citealp{thuan85}). 
By contrast, dwarf ellipticals (dEs) and dwarf spheroidals (dSphs)
have smoother visual appearances, mostly old stars, and low gas
content. 
dEs and dSphs, respectively, are the bright and faint representatives
of gas--poor dwarf galaxies in which star formation has ceased
(e.g., \citealp{bst84,kormendy85,bender92,binggeli94}).
In all of these types of dwarf galaxies, central surface brightness
increases with increasing luminosity\footnote{
Low--luminosity ellipticals or LLEs (e.g., M32) are structurally
similar to giant ellipticals and exhibit a trend of decreasing 
surface brightness with increasing luminosity.
}.
It seems natural to consider a possible evolutionary relationship
between dIs and dEs/dSphs
(e.g., \citealp{fl83,lf83,bmcm86,ds86,dp88,dp89,fb94}).
However, it is not yet well understood how dIs could have evolved or
transformed to become dEs/dSphs, as dIs have bluer colours and
lower surface brightnesses than luminous dEs/dSphs at a given
luminosity (e.g., \citealp{bmcm86,dp88}).
Moreover, while a relationship between iron abundance in stars and
galaxy luminosity has been observed in nearby dSphs
(\citealp{aaronson78,smith85,aaronson86,gg99,cote00,pb02}), 
\cite{rm95} and \cite{rms98} showed that oxygen abundances
for a number of field dSphs are about $+$0.3~dex higher than in field
dIs at similar luminosities and concluded that the predecessors
of these nearby dSphs could not have been dIs.

Owing to their relatively low gravitational potentials, dwarf galaxies
are thought to be susceptible to internal blow--out of their gaseous 
contents 
(e.g., \citealp{ds86,puche92,martin96,dellaceca97,strickland97}),
although some claim that blow--out signatures of dwarf galaxies have
not been definitively observed
(\citealp{bes94,skillman97,meurer98,mlf99,ft00}).
This motivates careful study of the chemical properties of field dIs,
which have the potential to uncover the relevance of gas flows to
evolution in isolation.
With knowledge about the history of field dIs, it becomes possible to
examine differentially the evolution of dIs in clusters, and thereby
pinpoint perturbations stemming from the different environments.
A subsequent comparison with the Virgo Cluster is the subject of
Paper~II \citep{lrm03}.

The outline of the paper is as follows.
The sample of field dwarfs is presented in \S~\ref{sec_sample}.
Observations and reductions are described in \S~\ref{sec_obs}.
The data analysis and a discussion of oxygen abundances are presented
in \S~\ref{sec_analysis} and \S~\ref{sec_oxyderive}, respectively.
The derived properties are listed in \S~\ref{sec_derivedprops}.
Two key diagrams are used as diagnostics of dI evolution: 
the oxygen abundance versus luminosity and the 
oxygen abundance versus baryonic gas fraction.
A discussion of these diagnostics and how the gas and stellar
constituents are used to define the gas fraction is presented 
in \S~\ref{sec_evolution}.
Finally, the conclusions are given in \S~\ref{sec_concl}. 

\section{The Sample of Dwarf Irregulars in the Field}
\label{sec_sample}

The sample of dIs in the field compiled by \cite{rm95} is the
starting point of research for this work.
The sample is defined by two criteria: 
(1) oxygen abundances are determined directly from measurements 
of the \othreea\ emission line, and 
(2) distances are determined from stellar properties following
the prescription in the appendix of \cite{rm95}.
Table~\ref{table_fieldsample} lists the field sample of 22 dwarf
irregular galaxies along with their basic properties.
A new addition to the sample is DDO~187 (\S~\ref{sec_ddo187}), which
also satisfies the above criteria.

To establish the best possible parameters for the field sample,
the literature was searched for measurements more recent than those
compiled by \cite{rm95}.
Direct oxygen abundances and well--determined distances were obtained
from data given in papers describing emission--line spectroscopy and
photometry of resolved stellar populations, respectively.
Tables~\ref{table_distupdate} and \ref{table_specupdate},
respectively, summarize the dwarf galaxies whose distances and
oxygen abundances have been updated. 
Of note also are updates to values of the \hi\ 21--cm flux 
\citep{hoffman96,stilisrael02} for GR~8, IC~1613, NGC~1569, 
and Sextans~B.

\subsection{Distances}

Distance measurements founded upon the observations of Cepheid
variable stars or the location of the tip of the red giant branch were
preferred. 
Distance determinations from Cepheids are founded upon the Cepheid
calibration by \cite{mf91}, whose value of the distance modulus to the
LMC was 18.50~mag. 
\cite{panagia99} measured the geometric expansion of the ring in the
supernova remnant SN1987A and derived a distance modulus of 18.58~mag.
This value of the distance modulus to the LMC is adopted for the
remainder of this work.  
Values of distance moduli from Cepheids are adjusted by $+0.08$~mag
to account for the revised distance to the LMC. 

\cite{lfm93} showed that the absolute magnitude in $I$ of the
tip of the red giant branch (TRGB), $M_I$, is relatively constant,
independent of metallicity and age, so long as the stars are
generally metal--poor (${\rm [Fe/H]} \la -0.7$ dex) 
and older than a few Gyr.
Thus, measurements of $M_I$ for resolved galaxies are good gauges 
of distance.
To account for modest variations in colour and metallicity in each
galaxy, absolute magnitudes in $I$ of the TRGB were obtained
following the prescription of \cite{rm95}. 
Distance moduli were derived from observed magnitudes in $I$,
corrected for extinction. 
In spite of the dispute regarding the range in distance moduli
to the LMC, it is reassuring that distances obtained
from TRGB measurements \citep{cioni00,sakai00} are consistent with and
independent of the distance derived from the SN1987A measurement.

For galaxies with more than one kind of observation, derived
distance moduli were averaged.
Generally, the errors in distance moduli obtained from either 
Cepheids or the TRGB are about 0.2 mag.
For the remaining galaxies in Table~\ref{table_fieldsample} with no
recent updates in the literature since 1995, distance moduli derived
by Richer \& McCall are adjusted by $+0.208$~mag, because the value
for the distance modulus to the LMC used by \cite{rm95} was
18.372~mag.

\subsection{Oxygen Abundances}

Many authors have reported emission--line spectra of \hii\ regions in
field dIs since \cite{rm95} compiled results from the literature. 
Oxygen abundances have been computed from these spectra using the
published measurements of \otwo, \othreea, and \othree\ emission lines.
These updates are listed in Table~\ref{table_specupdate}, which
also incorporates new spectroscopic data for five field dIs obtained
at the Steward Observatory and at the Observatorio Astron\'omico
Nacional (OAN) in M\'exico.
The new data are described in \S~\ref{sec_obs}.

\subsection{A New Addition: DDO 187}
\label{sec_ddo187}

DDO~187 (UGC~9128) is an isolated late--type, gas--rich dwarf galaxy,
which appears to have disk-halo structure.
The disk stellar component extends much farther out than the gas
component; the outer component does not contain young stars 
\citep{atk00}. 
A distance based on the measurement of two Cepheids has been
determined by \cite{hoessel98b}.
We adopt here the result of \cite{atk00}, whose TRGB distance is
almost three times smaller than the Cepheid distance.

H~II regions at the centre have been observed by
\cite{strobel91}, \cite{hhg93}, and \cite{vanzee97}. 
\cite{vanzee97} obtained spectroscopy for the two 
brightest \hii\ regions; \othreea\ was detected and an oxygen
abundance was determined for DDO~187--1, which is the brighter of the
two \hii\ regions.
The published fluxes for DDO~187--1 were reanalyzed in a manner
similar to the analysis of the five field dIs for which new spectra
were acquired. 

\section{New Spectra: Observations and Reductions}
\label{sec_obs}		

\begin{figure}
\epsscale{0.6}
\plotone{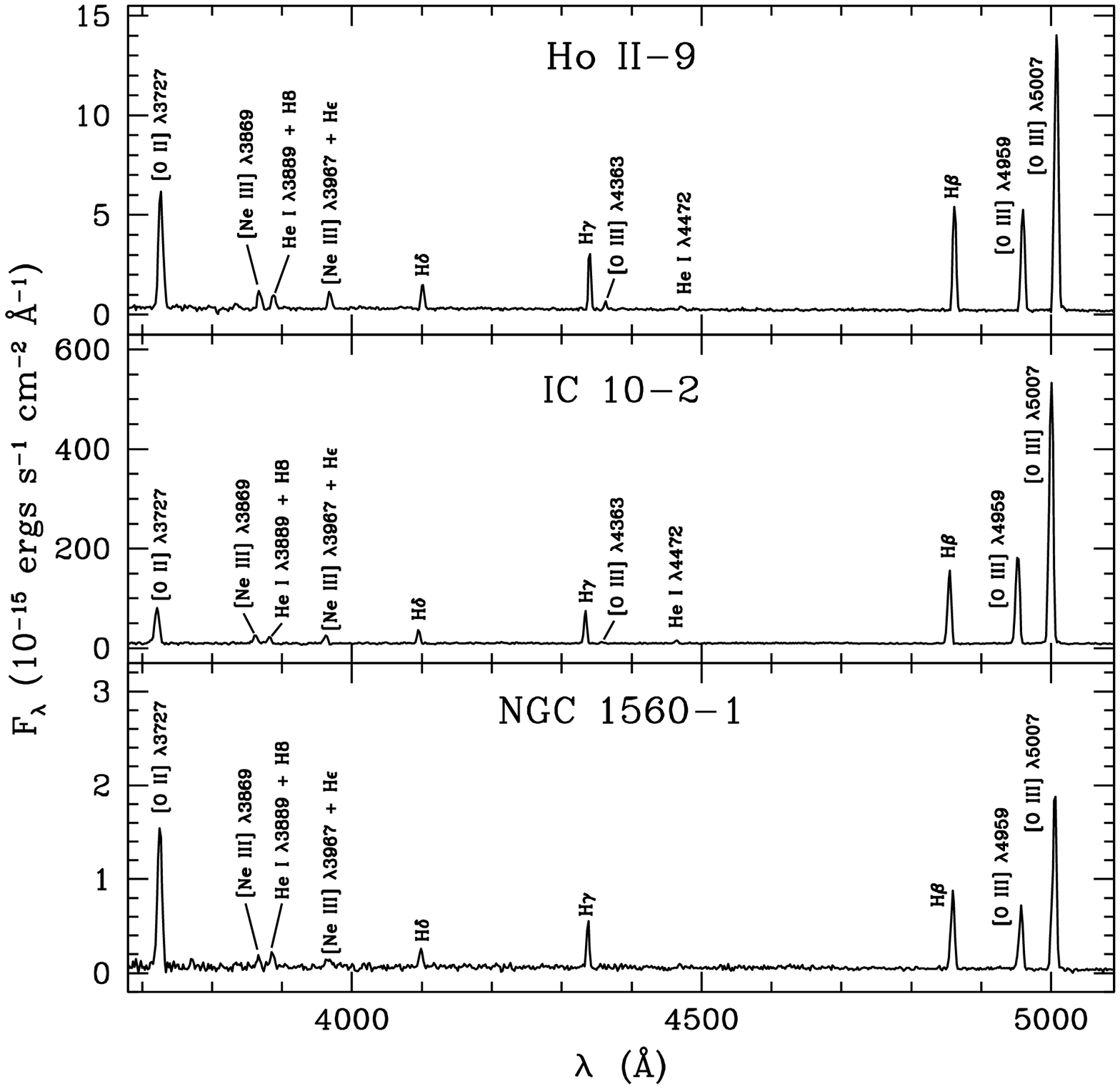}
\caption{
Optical emission--line spectra of Holmberg II--9, IC~10--2, 
and NGC~1560--1 from Steward Observatory.
The observed flux per unit wavelength is plotted versus wavelength.
Key emission lines are labelled.
}
\label{fig_fieldspc1}
\end{figure}

\begin{figure}
\epsscale{0.6}
\plotone{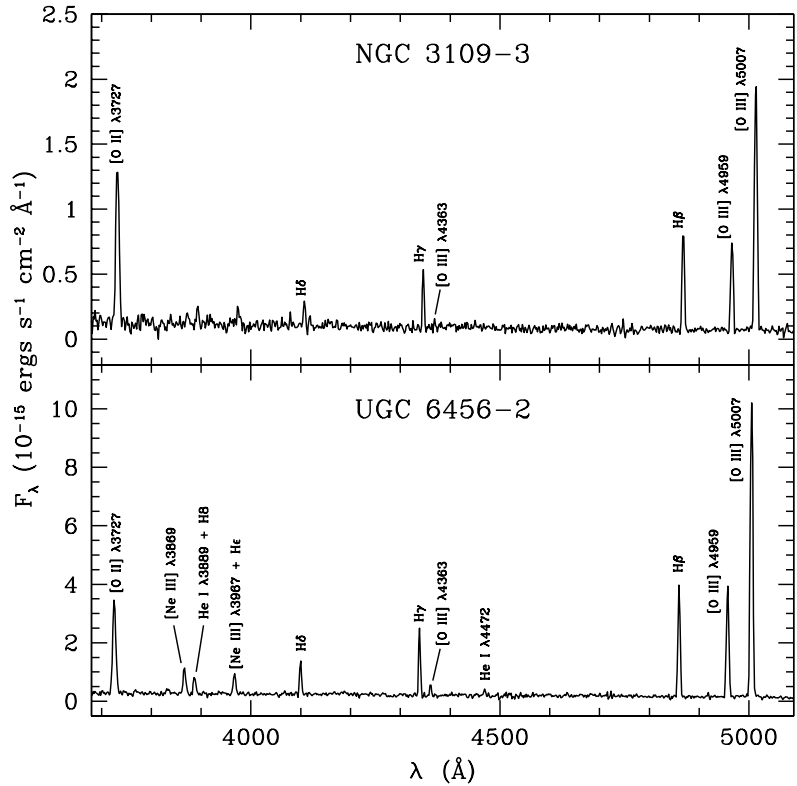}
\caption{
Optical emission--line spectra of NGC~3109--3 and UGC~6456--2
from Steward Observatory.
The observed flux per unit wavelength is plotted versus wavelength.
Key emission lines are labelled.
}
\label{fig_fieldspc2}
\end{figure}

A program of long--slit spectroscopy was carried out at the Steward
Observatory and the OAN to acquire optical spectra for a number of
dwarf galaxies in the field.
The primary goal was to obtain sufficiently long exposures to detect
the temperature--sensitive \othreea\ emission line.
Results for five dIs are presented here.
In each of the dIs, \hii\ regions for which spectra were obtained were
matched with identifications in the published literature; 
these are described below.
Spectra for \hii\ regions Holmberg~II--9, IC~10--2, NGC~1560--1,
NGC~3109--3, and UGC~6456--2 are shown in Figures~\ref{fig_fieldspc1}
and \ref{fig_fieldspc2}.
Details of the instrumentation employed and the log of observations
are listed in Tables~\ref{table_stewardspm} and \ref{table_obslog},
respectively. 

Spectra were reduced in the standard manner using 
IRAF\footnote{IRAF is distributed
by the National Optical Astronomical Observatories, which is operated
by the Associated Universities for Research in Astronomy, Inc., under
contract to the National Science Foundation.}
routines from the long--slit spectroscopy reduction package
``specred.''
Dome flat exposures were used to remove pixel--to--pixel variations 
in response. 
Cosmic rays were identified and deleted manually.
Final one--dimensional flux--calibrated spectra for each \hii\ region
were obtained via unweighted summed extractions. 

\subsection{H II Regions in Holmberg II}

Several of the \hii\ regions in Holmberg~II were spanned by the long
slit employed at Steward.
Table~\ref{table_holm2_h2rs} lists the \hii\ regions for which
spectra were acquired and in which the \othreea\ line was detected.
Locations of the \hii\ regions were determined by comparing
finding charts used by M. McCall against published \halpha\ data
\citep{hg85,hsk94}.

\subsection{H II Regions in IC 10}
\label{sec_ic10_h2rs}

Table~\ref{table_ic10_h2rs} lists the \hii\ regions for which
spectra were acquired at Steward.
These same data have been used previously to study the extinction
in IC~10 \citep{richer01}.
By inspection of finding charts provided by M. McCall, the
locations of the observed \hii\ regions were matched with \hii\
regions identified by \cite{lequeux79} and \cite{hl90}.

In addition to the Steward data, long--slit spectra between 3450~\AA\
and 7450~\AA\ were obtained with the 2.1--metre telescope at the OAN
on 1994 December~1 (UT).
Spectra were acquired specifically to measure \halpha\ and \hbeta\ to
provide a larger wavelength baseline for a truer estimation
of the reddening.
The spectra were reduced in a manner similar to the Steward data.
\hii\ regions identified at the OAN were matched with those identified
at Steward, so that the appropriate reddening corrections could be
applied directly to each \hii\ region.

\subsection{H II Regions in NGC 1560}

The long slit was placed lengthwise along the major axis
of the galaxy to obtain spectra of \hii\ regions located in the disk.
Table~\ref{table_n1560_h2rs} lists the \hii\ regions for which spectra
were acquired.
From inspection of finding charts, \halpha\ images provided by
M. McCall, and the broadband images by \cite{lm93}, 
approximate locations of observed \hii\ regions with respect to
the measured stars in \cite{lm93} were determined.
The \hii\ region NGC 1560--1 is visible as the brightest
``concentration'' located near the centre of the galaxy in the
$V$--band image of \cite{lm93}. 
For \hii\ regions NGC~1560--6 and NGC~1560--7, \othreea\ was
not detected because of the poor quality of the spectra. 

\subsection{H II Regions in NGC 3109}

Table~\ref{table_n3109_h2rs} lists the \hii\ regions for which
spectra were measured.
By inspection of the finding charts by M. McCall, the locations of the
observed \hii\ regions were matched with \cite{rm92}.
A continuum--subtracted \halpha\ image may also be found in 
\citet[Figure 11]{hhg93}, but it is difficult to match the \hii\
regions in their \halpha\ image with the \hii\ regions
identified in the \othreec\ image by \cite{rm92}.

\subsection{H II Regions in UGC 6456}

The long slit was placed on the galaxy in an orientation similar to
that employed by \cite{tully81}. 
Table~\ref{table_u6456_h2rs} lists the \hii\ regions for which spectra
were measured.
By inspection of the finding charts by M. McCall, the locations of the
observed \hii\ regions were matched with \cite{tully81} and
\cite{lynds98}.

\section{Measurements and Analysis}
\label{sec_analysis}

Emission--line strengths were measured using locally--developed
software.
Flux ratios were corrected for underlying Balmer absorption
with an equivalent width 2~\AA\ \citep{mrs85}.
Corrections and analyses were performed with SNAP
(Spreadsheet Nebular Analysis Package, \citealp{snap97}). 

A reddening, $E(B-V)$, was computed from Balmer flux ratios using 
\begin{equation}
\log\frac{I(\lambda)}{I({\rm H}\beta)} = 
\log\frac{F(\lambda)}{F({\rm H}\beta)} + 
0.4\,E(B-V)\,\left[A_1(\lambda) - A_1({\rm H}\beta)\right].
\label{eqn_reddening}
\end{equation}
$F(\lambda)/F(\hbeta)$ and $I(\lambda)/I(\hbeta)$ are the
observed flux ratio and corrected intensity ratio, respectively, at
wavelength $\lambda$.
$A_1(\lambda)$ is the extinction in magnitudes for $E(B-V) = 1$, 
i.e., $A_1(\lambda) = A(\lambda)/E(B-V)$, where 
$A(\lambda)$ is the monochromatic extinction in magnitudes.
Values of $A_1$ were obtained from the \cite{cardelli89}
reddening law as defined by a ratio of the total to selective
extinction, $R_V$, equal to 3.07, which in the limit of zero reddening
is the value for an A0V star (e.g., Vega) with intrinsic $B-V$ colour
equal to zero.
For \halpha-- and \hgamma--based reddenings, the ratios 
$I$(\halpha)/$I$(\hbeta) = 2.86 and
$I$(\hgamma)/$I$(\hbeta) = 0.468 were adopted, respectively,
which are appropriate for typical conditions within \hii\ regions 
($T_e$ = 10$^4$~K and $n_e$ = 100~cm$^{-3}$; \citealp{osterbrock}).

After correcting line ratios for the initial estimate of the
reddening, the temperature was estimated and a second value of the
reddening was computed. 
SNAP was used to compute temperatures by directly solving the
equations of statistical equilibrium and determining the emissivities
for collisionally excited lines of O$^+$ and O$^{+2}$ using the
five--level atom approximation (see, e.g., \citealp{fivel}). 
In the absence of \othreea, the electron temperature was
assumed to be $T_e = 10^4$~K.
Because the spectra acquired at Steward were obtained at wavelengths
between 3600~\AA\ and 5100~\AA, the lines \halpha, \ntwo, and \stwo\ 
were not detected. 
As the density--dependent line ratio, $I$(\stwoa)/$I$(\stwob),
could not be formed, a value of $n_e$ = 100~cm$^{-3}$ was adopted.
It was found that the reddening values did not change significantly
from the first to the second iteration.
Also, the derived temperature in the second iteration was found to
be very similar to the value from the first iteration.

H$\gamma$--based reddenings were used to correct spectra of \hii\
regions in Holmberg~II, NGC~3109, and UGC~6456.
\halpha--based reddenings were used to correct spectra of \hii\
regions in IC~10 and NGC~1560.
$F$(\hgamma)/$F$(\hbeta) was not used in a number of spectra
(e.g., IC~10, NGC~1560) because of severe absorption or poor signal
at \hgamma. 

Observed and corrected line ratios for the five field dIs observed
at Steward and at the OAN are listed in 
Tables~\ref{table_holm2_data} to \ref{table_u6456_data} inclusive.
The listed errors for the observed flux ratios at each wavelength
$\lambda$ account for the errors in the fits to the line profiles,
their surrounding continua, and the relative error in the sensitivity
function stated in Table~\ref{table_obslog}.  
The error in the \hbeta\ reference line is not included in the
observed ratios.
The uncertainty in the correction for underlying Balmer absorption
was taken to be zero.
Errors in the corrected intensity ratios account for errors in the 
specified line and in the \hbeta\ reference line.
Brief comments about the line ratios for each dI are presented below.

\subsection{Holmberg II}

Observed flux and corrected intensity ratios are listed 
in Table~\ref{table_holm2_data}.
No variations in $T_e$ were observed to within the computed errors.
The reddening values for Holmberg~II were found to be small and
consistent with zero.
Thus, zero reddening was adopted and the observed flux ratios were
subsequently compensated only for underlying Balmer absorption with
an equivalent width of 2~\AA.

\subsection{IC 10}


Observed flux and corrected intensity ratios obtained at the OAN are
listed in Table~\ref{table_ic10_oan}.
Derived reddenings for each \hii\ region are also listed.
As stated in \S~\ref{sec_ic10_h2rs}, the OAN spectra were obtained for
the purpose of deriving a reddening value from \halpha\ and
\hbeta\ fluxes.
For the Steward data, intensity ratios were derived from flux ratios by
compensating for underlying Balmer absorption with an equivalent
width of 2~\AA\ and for the reddening derived for each \hii\ region
from the OAN data.
Observed flux and corrected intensity ratios are listed in
Table~\ref{table_ic10_data}.
Errors in the corrected intensity ratios account for errors
in the flux at the specified line, errors at the \hbeta\ reference
line, and errors in the reddening values from data obtained at
the OAN.

Line intensity ratios for a number of \hii\ regions IC~10 have also
been reported by \cite{richer01}. 
The flux ratios for IC~10--2 and IC 10--4 (Table~\ref{table_ic10_data})
are comparable to those reported for HL111c and HL111b, respectively,
which were derived independently from the same set of observations.

\subsection{NGC 1560}

Because NGC~1560 is at a low galactic latitude, there is expected to
be non--negligible foreground extinction along the line--of--sight
\citep{buta_mccall99}. 
A second observing program at Steward was designed specifically to
measure \halpha\ and \hbeta\ fluxes, because the \hgamma\ fluxes were 
found to be unusable.
The long slit was placed on the northeast and southwest
regions of the galaxy, so the \hii\ regions observed were different
from those examined at blue wavelengths. 
Reddening values were derived from $F$(\halpha)/$F$(\hbeta).
An average reddening of $E(B-V) = +0.36$~mag was computed from the
\hii\ regions 1~NE, 1~SW, 2~SW, 3~SW, and 6~SW.
This value was adopted to correct line ratios for \hii\ regions
observed in the blue.
Line data and the reddenings are listed in
Table~\ref{table_n1560_reddata}. 

Observed flux and corrected intensity ratios for the blue
spectroscopic data are listed in Table~\ref{table_n1560_data}.
Flux ratios for the two \hii\ region spectra NGC~1560--1 and
NGC~1560--2 are in agreement with those determined independently
by M. G. Richer (2000, private communication) from the same set of
observations.

\subsection{NGC 3109}

Observed flux and corrected intensity ratios are presented in
Table~\ref{table_n3109_data}.
Reddening values derived from $F$(\hgamma)/$F$(\hbeta) were found to
be near zero, consistent with the small value listed in NED.
Zero reddening was adopted and the observed flux ratios were
subsequently compensated only for underlying Balmer absorption with
an equivalent width of 2~\AA. 

\subsection{UGC 6456}

Observed flux and corrected intensity ratios are listed 
in Table~\ref{table_u6456_data}.
Values of the reddening derived from $F$(\hgamma)/$F$(\hbeta)
were consistent with zero. 
Zero reddening was adopted and the observed flux ratios were
subsequently compensated only for underlying Balmer absorption with
an equivalent width of 2~\AA.

\subsection{DDO 187}

The fluxes reported by \cite{vanzee97} were reanalyzed in the same
manner as the five previous dIs.
Zero reddening was assumed and the observed flux ratios
were corrected for underlying Balmer absorption with an 
equivalent width of 2~\AA. 
Observed and corrected line ratios are listed in 
Table~\ref{table_ddo187_data}.

\section{Oxygen Abundances}
\label{sec_oxyderive}

The direct or standard method of obtaining oxygen abundances 
from emission lines is applicable to any object with detectable
\othreea\ and for which the doubly ionized O$^{+2}$ ion is the
dominant form of oxygen \citep{osterbrock}. 
The method by which oxygen abundances are derived with the
standard method is summarized in \cite{dinerstein90}.
Computations for the present work were performed with SNAP.

Given a temperature and a density, an oxygen abundance is determined
from strong emission arising from singly-- and doubly--ionized oxygen.
At temperature $T_e$, the relative abundances of singly-- and
doubly--ionized oxygen by number are, respectively,
\begin{eqnarray}
\frac{N({\rm O}^+)}{N({\rm H})}
& = & \frac{I({\rm [O\;II]}\lambda\,3727)}{I({\rm H}\beta)} \cdot 
\frac{j({\rm H}\beta;\;n_e,\,T_e)}
{j({\rm [O\;II]}\lambda\,3727;\;n_e,\,T_e)}, \\
\frac{N({\rm O}^{+2})}{N({\rm H})} 
& = & \frac{I({\rm [O\;III]}\lambda\,5007)}{I({\rm H}\beta)} \cdot 
        \frac{j({\rm H}\beta;\;n_e,\,T_e)}
        {j({\rm [O\;III]}\lambda\,5007;\;n_e,\,T_e)}, 
\label{eqn_computeabund}
\end{eqnarray}
where $I$ is the intensity of the line at wavelength $\lambda$,
$j$ is the volume emissivity of the line
at density $n_e$ and temperature $T_e$, and 
$N(A^{+k})$ is the abundance by number of the atomic species
in the $k$th--ionized state responsible for the line.
Data for Balmer line emissivities were taken from \cite{sh95}.
Emissivities for neutral and ionic oxygen were computed using
spontaneous emission coefficients from \cite{wiese96} 
and collision strengths from various sources
\citep{pradhan76,mb93,lb94,bk95}.
The total oxygen abundance by number, $N$(O)/$N$(H), is obtained from
the sum 
\begin{equation}        
\frac{N({\rm O})}{N({\rm H})} = f \cdot 
\left[ \frac{N({\rm O}^+)}{N({\rm H}^+)} + 
\frac{N({\rm O}^{+2})}{N({\rm H}^+)} \right] ,
\label{eqn_totaloxyabund}
\end{equation}
where $f$ is an ionization correction factor to account for 
unobserved oxygen ions. 
Since there was little or no He~II emission, the ionization correction
factor could be assumed to be unity.

\section{Derived Properties}
\label{sec_derivedprops}

Derived properties for the five dIs observed at Steward and OAN are
listed in Table~\ref{table_derivedprops}.
For at least one \hii\ region in each of Holmberg~II, IC~10, NGC~3109,
and UGC~6456, \othreea\ was detected and an oxygen abundance was
derived directly.
A meaningful lower limit to the oxygen abundance was obtained for
NGC~1560. 
The listed properties include \hbeta\ intensities corrected for
underlying Balmer absorption and reddening, derived and adopted values
of the reddening, \hbeta\ emission equivalent widths corrected for
underlying Balmer absorption, O$^{+2}$ electron temperatures, and
resulting oxygen abundances. 
Errors in oxygen abundances were computed from the maximum and
minimum possible values, given the errors in the reddening,
temperature, and line intensities.

\subsection{Holmberg II}

Oxygen abundances for \hii\ regions Ho~II--5, Ho~II--8, and
Ho~II--9 were adopted, because these \hii\ regions are independent of
each other and have the largest \hbeta\ intensities.
The abundance for Ho~II--7 was not adopted, because the corresponding 
spectrum encompasses the spectra of both Ho~II--6 and Ho~II--8 (see
Table~\ref{table_holm2_h2rs}). 
Oxygen abundances for three \hii\ regions reported by
\cite{mmdo91} were re--analyzed in the same manner as the 
Steward data. 
The average oxygen abundance listed in Table~\ref{table_fieldsample}
was computed using results for the three selected \hii\ regions in 
the present analysis and results for the three \hii\ regions in
\cite{mmdo91}. 

\subsection{IC 10}

Oxygen abundances for \hii\ regions IC~10--2 and IC~10--3 were
adopted.
The abundance for IC~10--1 was not adopted, because the extracted
spectrum encompasses the spectra for IC~10--2 and IC~10--3 (see also
Table~\ref{table_ic10_h2rs}). 
The oxygen abundance for IC~10 listed in Table~\ref{table_fieldsample}
is an average of abundances for IC~10--2 and IC~10--3.
Computed errors in derived oxygen abundances include errors
in both reddenings and temperatures.
For IC~10--2, the derived oxygen abundance 
(12$+$log(O/H) = $8.32 \pm 0.14$) agrees with the value independently
derived by \cite{richer01} for the same Steward data, but is just
outside of the value published by Lequeux et al.
\citeyearpar[their \hii\#1;][]{lequeux79}.

\subsection{NGC 1560}

The \othreea\ line was not detected.
To determine a lower limit to the oxygen abundance for the galaxy,
two spectra with the highest signal--to--noise were selected from the
seven \hii\ regions observed.
For \hii\ regions NGC~1560--1 and NGC~1560--2, the lower limits are
12$+$log(O/H) $\geq 8.05$ and $\geq 7.89$, respectively. 
The mean of lower limits is a lower limit to the mean, provided that
each lower limit is lower than the true abundance values \citep{rm95}.
So, the mean of the lower limits obtained from the present analysis was
12$+$log(O/H) $\geq 7.97$, which was adopted as the lower limit 
to the oxygen abundance for NGC~1560.

\subsection{NGC 3109}

In \hii\ region NGC~3109--3, \othreea\ was detected, from which an
oxygen abundance of 12+log(O/H) = $7.73\,\pm\,0.33$ was derived.
The oxygen abundance for NGC~3109 listed in \cite{rm95}, which
was derived from the same Steward data, is $1\sigma$ higher than the
present value. 
The ratio $F$(\othreea)/$F$(\hbeta) for NGC~3109--3 is
$(5.73\,\pm\,2.04)\,$\%; the measurement has an error of 36\%. 
It is worth noting that the value of the oxygen abundance given in
\cite{rm95} is based upon a previous coarse analysis of the same
Steward data in which \othreea\ was not measured with a high degree of 
confidence.
From M. G. Richer (2000, private communication), their
$F$(\othreea)/$F$(\hbeta) was $(3.13\,\pm\,1.49)\,$\% with 
a measurement error of 48\%. 
The improvement here was due to an improvement in the procedure 
used to extract the \hii\ region spectrum.
The new value of the oxygen abundance was adopted for NGC~3109 and
listed in Table~\ref{table_fieldsample}. 
The error in the adopted oxygen abundance is large (0.33~dex), because
of significant errors ($\simeq$~10\%) in each of the corrected
intensity ratios for [O~II] and [O~III], as well as an error near 40\%
in the corrected intensity ratio for \othreea.

\subsection{UGC 6456}

The oxygen abundance for the brighter \hii\ region UGC~6456--2 
(12$+$log(O/H) = 7.45) was adopted for the present work.
\cite{itl97}, \cite{lynds98}, and \cite{hh99} provide 
more recent measurements.
A homogeneous treatment of these data gives oxygen abundances
of 12$+$log(O/H) = 7.65, 7.74, and 7.73, respectively.
The oxygen abundance for UGC~6456 listed in
Table~\ref{table_fieldsample} is an average of the abundance derived
here and abundances from the three measurements in the literature. 

\subsection{DDO 187}

For DDO 187, the derived values for the electron temperature and oxygen
abundance agree with the results in \cite{vanzee97}. 
From the present analysis, an oxygen abundance of 12$+$log(O/H) =
$7.69 \pm 0.09$ is derived for DDO~187--1.

\section{The Evolution of dIs in the Field}
\label{sec_evolution}

An understanding of the properties and their correlations for a
sample of dIs in the relatively low density environment of the 
field is required to explore and discern possible differences for
galaxy evolution in high density environments.
The metallicity--luminosity (O/H vs. $M_B$) and the metallicity--gas
fraction (O/H vs. $\mu$) diagrams are key diagnostics of 
evolution. 

\subsection{Fitting Procedure}
\label{sec_fitting}

A best--fit line for the correlation between two parameters with
comparable errors is obtained with the geometric mean functional
relationship \citep{regression_book}, which assumes similar
dispersions in both observables.
The geometric mean functional relationship relies upon the
minimization of the sum of areas bounded by the shortest horizontal
and vertical lines from each data point to the best--fit line. 
While maximum likelihood techniques are more appealing for best
results, these methods do not work as well when the datasets are small
\citep{regression_book}.
Here, typical errors are assumed to be about 0.1 to 0.2~dex for
both dependent and independent variables.

For the desired relation $Y$ versus $X$, the fit is described by
\begin{equation}
Y = y_0 + m X.
\label{eqn_geommean_fit}
\end{equation}
Initially, two linear least--squares fits are obtained : 
$Y = b_0 + b_1\, X$ and $X = a_0 + a_1\, Y$.
The desired slope, $m$, is obtained from the geometric mean of the
slopes from the two linear least--squares fits
\begin{equation}
m = \left( \frac{b_1}{a_1} \right)^{1/2}.
\label{eqn_geommean_slope}
\end{equation}
The desired intercept, $y_0$, is given by
\begin{equation}
y_0 = \langle Y \rangle - m\,\langle X \rangle,
\label{eqn_geommean_inter}
\end{equation}
where $\langle X \rangle$ and $\langle Y \rangle$ are the averages of 
$X$ and $Y$ values, respectively.
For a given fit, all points are equally weighted.

\subsection{Oxygen Abundance versus Luminosity}
\label{sec_oxylum}

The metallicity--luminosity diagram has long been considered to be
representative of a metallicity--mass relationship for dIs.
\cite{skh89} and \cite{rm95} showed that oxygen abundances 
in dIs increase with increasing galaxy luminosity in $B$.
\cite{hgo98} claimed from their sample of dIs that this relationship
is weaker than previously thought, but low signal--to--noise in their
\othreea\ measurements could account for their lack of an observed
relation \citep{pilyugin01}.
\cite{kiss02} obtained a metallicity--luminosity correlation for a
sample of emission line objects from their KPNO International
Spectroscopic Survey. 
However, their sample included objects at higher luminosities
($M_B \simeq -22$), which may exhibit abundance gradients.
Only a small minority of their objects were detected at \othreea\
and other estimates using empirical or bright--line methods 
(e.g., \citealp{mcgaugh91,pilyugin00})
were required to derive the remaining oxygen abundances.

\begin{figure}
\epsscale{0.5}
\plotone{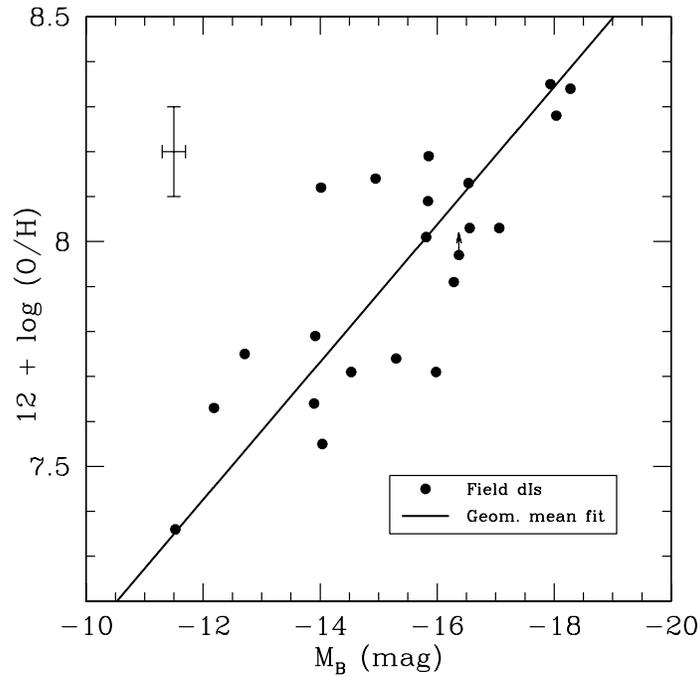}
\caption{
Oxygen abundance versus absolute magnitude in $B$ for field dIs.
Galaxy luminosity increases to the right.
The filled circles mark the field dIs. 
The arrow indicates a lower limit to the oxygen abundance
for NGC~1560.
Oxygen abundances for the field dwarfs were determined
directly from \othreea\ measurements.
The solid line is a fit to the field dIs 
(Equation~\ref{eqn_zmb_new}) using the geometric mean functional
relation \citep{regression_book}. 
The error bars indicate typical uncertainties in the oxygen abundance
and absolute magnitude.
The uncertainty is at most 0.1~dex for oxygen abundances which
were determined directly from \othreea\ measurements.
An uncertainty of 0.2~mag in absolute magnitude is typical of what can
arise from distance determinations.
}
\label{fig_zmb_field}
\end{figure}

Based upon the data presented here, the fitting method described in
\S~\ref{sec_fitting} was applied to obtain a revised fit to the
metallicity--luminosity relation. 
A good correlation is seen in a plot of oxygen abundance against
luminosity for the sample of field dIs as shown in
Figure~\ref{fig_zmb_field}.
The relation for the sample of field dIs is 
\begin{equation}
12 + \log{\rm (O/H)} = (5.59 \pm 0.54) + (-0.153 \pm 0.025)\,M_B
\label{eqn_zmb_new}
\end{equation}
and the root--mean--square (rms) in log(O/H) is $\sigma = 0.175$~dex.
This fit is shown as a solid line in Figure~\ref{fig_zmb_field}.
Equation~(\ref{eqn_zmb_new}) is consistent with the relation
determined by \cite{rm95} for dwarfs brighter than $M_B = -15$.
Combining newly acquired data with updates to distances and abundances
to other dwarfs has not significantly altered this relation,
although the scatter at low luminosities has been reduced
(compare with \citealp[Figure~4,][]{rm95}).
Equation~(\ref{eqn_zmb_new}) will be adopted as the
metallicity--luminosity relation for the sample of field dIs.

\subsection{Gas, Stellar, and Baryonic Masses} 

The fraction of baryons in gaseous form is a fundamental quantity,
because it determines the metallicity within the ``closed box'' model
of chemical evolution.
Dynamical masses have been used in the past to gauge the gas fraction
(e.g., \citealp{lequeux79,mc83,pagel86}).
However, a difficulty with the dynamical mass is that it may be
dominated by non--baryonic dark matter, which does not participate in
nucleosynthesis. 
Moreover, dynamical masses are notoriously difficult to measure in dIs,
because random or turbulent motions can dominate over ordered or
rotational motions, especially at low luminosities
(see, e.g., \citealp{lsy93}). 
Instead, it is more prudent to evaluate the gas fraction by estimating
masses for the luminous components.

The gas in dIs is the raw material out of which stars and metals are
formed and consists mostly of hydrogen and helium.
The largest constituent of gas in dIs is assumed to be neutral atomic
hydrogen. 
The composition of gas in molecular form within dIs remains mostly an
unknown quantity (see, e.g., \citealp{ys91,hs93,israel95}).
Although CO is used to trace the molecular hydrogen content in galaxies, 
the conversion from CO to H$_2$ masses is uncertain, because there 
is much debate about the ``universality'' of the conversion factor.
Fortunately, molecular gas is not expected to contribute greatly to 
the total gas mass in dwarf galaxies, owing to low metallicities and
low dust--to--gas ratios \citep{ys91,lf98,vidal00}.
Finally, ionized gas contributes negligibly to the mass \citep{sme01}. 

The \hi\ mass in solar masses is given by the following
equation \citep{roberts75,rh94}:
\begin{equation}
M_{H I} = 2.356 \times 10^5 \: F_{21} \: D^2,
\label{eqn_h1mass}
\end{equation}
where $F_{21}$ is the 21--cm flux integral in Jy~km~s$^{-1}$ and 
$D$ is the distance in Mpc.
Accounting for helium and other metals, the total gas mass 
in solar masses is given by
\begin{equation}
M_{\rm gas} = M_{H I}/X,
\label{eqn_gasmass}
\end{equation}
where $X$ is the fraction of the gas mass in the form of hydrogen,
adopted to be 0.733.

The mass in stars, $M_{\ast}$, is often computed from the product of
the blue luminosity, $L_B$, with an assumed constant stellar
mass--to--light ratio, $M_{\ast}/L_B$. 
However, this does not account for the possible contamination of the
luminosity by bright star--forming regions, which may introduce
significant variations in $M_{\ast}/L_B$ from galaxy to galaxy. 
The problem may be particularly severe for dIs, because a star
formation event of a given mass would have a proportionately larger
effect on the light than it does in a giant spiral galaxy.
To account for varying rates of star formation, computing the stellar
mass requires a synthesis of stellar populations and it would be ideal
to know the flux contributions by old and young stars across a large
range of wavelengths.
While some recent syntheses of stellar populations within dIs have
been founded upon broadband optical and near--infrared colours
(e.g., \citealp{krueger91,kfa94}), all that is widely available for
colours of dIs in the present work is $B-V$. 
A method is briefly described below which yields an estimate for
$M_{\ast}/L_B$ tailored to the particular mixture of young and old
stars in a dI. 
Originally developed by M. L. McCall as part of a
long--term investigation of the masses of galaxies, this method
is described more fully in \cite{lee01}.

Stellar masses for field dIs were derived by supposing that dIs
consist only of a ``young'' disk component and an ``old'' disk
component.
The two--component method is founded upon the assumption that the
luminosity of young stars created in a recent burst or bursts of star
formation does not overwhelm that from the old stars.
Such is the case for dIs, but not for BCDs \citep{papaderos96,pt96}.
The method is exact if a burst is superposed upon an old population.
A one--component formalism to compute the stellar mass from a
``typical'' $M_{\ast}/L_B$ is not as good as the two--component
version, because the stellar mass--to--light ratio may depend on
luminosity, and may even vary greatly from galaxy to galaxy at a given
luminosity. 
\cite{lee01} has shown that the two--component method gives a
tighter correlation between oxygen abundance and the gas mass fraction
than the one--component method; see also the discussion at the end of
\S~\ref{sec_oxymu}. 

The mass of stars is given by the sum of the mass in the young
and old components
\begin{eqnarray}
M_{\ast} & = & M_{{\ast},{\rm yng}} + M_{{\ast},{\rm old}} \nonumber \\
& = & \left( \frac{M_{\ast}}{L_B} \right)_{\rm yng} \, L_{B,{\rm yng}}
    + \left( \frac{M_{\ast}}{L_B} \right)_{\rm old} \, L_{B,{\rm old}},
\label{eqn_mstartotal}
\end{eqnarray}
where $(M_{\ast}/L_B)_{\rm yng}$ and $(M_{\ast}/L_B)_{\rm old}$
represent the stellar mass--to--light ratios for the young and old
components, respectively, and 
$L_{B,{\rm yng}}$ and $L_{B,{\rm old}}$ represent 
the luminosity contributions in $B$ from the young and old 
components, respectively.
The luminosity contributions in $B$ from the young and old 
components are written as
\begin{eqnarray}
L_{B,{\rm yng}} & = & (1 - f_{\rm old}) \, L_B, \\
L_{B,{\rm old}} & = & f_{\rm old} \, L_B.
\label{eqn_lum_yng_old}
\end{eqnarray}
The fraction of light in $B$ contributed by old stars,
$f_{\rm old}$, is given by
\begin{equation}
f_{\rm old} =
\frac{10^{-0.4 (c_{\rm yng} - c_d)} - 1}
        {10^{-0.4 (c_{\rm yng} - c_{\rm old})} - 1},
\label{eqn_fold}
\end{equation}
where $c_{\rm yng}$, $c_{\rm old}$, and $c_d$ are,
respectively, the $B-V$ colours of the young stars, the old stars,
and the entire dI.

The young stellar component is presumed to have properties similar to
that of I~Zw~18, whose light is known to be dominated by young stars
with ages less than 10~Myr old.
The intrinsic $B-V$ colour assigned to the young stellar population,
\begin{equation}
c_{\rm yng} = (B-V)^0_{\rm yng} = -0.03\;{\rm mag},
\label{eqn_bv_yng}
\end{equation} 
is that observed for young dwarf stars ($M_V \la +4$) in the solar
neighbourhood 
\citep{bs84,vanderkruit86} 
and is consistent with the colour observed for I~Zw~18
\citep{legrand00}. 
The mass--to--light ratio in $B$ for young stars is
\begin{equation}
(M_{\ast}/{L_B})_{\rm yng} = 0.153\;M_{\odot}/L_{B,{\odot}}.
\label{eqn_mstarlb_yng}
\end{equation}
This mass--to--light ratio comes from models of the distribution of
young massive main--sequence stars ($M_V \leq +4.2$) perpendicular to
the Milky Way disk, which employ local measures of the luminosity
function and mass--luminosity relation \citep{bahcall84,bs84}. 
The mass--to--light ratio for young stars is taken to be constant,
because the initial mass function is found not to vary significantly
from galaxy to galaxy
(e.g., \citealp{massey95a,massey95b,holtzman97}).
The precise value is not of critical importance, because the mass in
typical dwarf galaxies is dominated by old stars and gas.

The ``old'' component is assumed to consist of old stars and
whatever baryonic dark matter is disk--like in its distribution.
The intrinsic $B-V$ colour assigned to the old stellar population
is given by
%
%
\begin{eqnarray}             
\lefteqn{
c_{\rm old} = (B-V)^0_{\rm old} = (-0.0423 \pm 0.0031) \cdot
}
\nonumber \\
& & \left[ M_{B,{\rm old}} + 1.0454\, \langle \mu^0_{B,{\rm eff}}
   \rangle_{\rm old} \right] + (1.177 \pm 0.019),
\label{eqn_bv_old}
\end{eqnarray} 
where $M_{B,{\rm old}}$ is the absolute magnitude in $B$ of
the old disk, and
$\langle \mu^0_{B,{\rm eff}} \rangle_{\rm old}$ is an estimate for the
dust--free mean surface brightness in $B$ of the old disk that would
be observed within a circular aperture with radius equal to the
effective radius of the disk, if the disk were an oblate spheroid seen
face--on.
The colour is estimated by assuming that the relationship between
$B-V$, luminosity, and surface brightness in old stars is similar to
that observed for the ensemble of old stars in disk--like or
exponential systems (e.g., dSphs) where star formation has stopped or
is occurring at a very low rate, and for the old disk of the Milky Way.
Dwarf elliptical galaxies in the Local Group and in the Fornax
cluster are used to set the relationship; dwarfs as faint
as $M_B \approx -9$ set the low--luminosity end of the range.
The disks of the Virgo dwarf galaxy VCC~1448, the low surface
brightness spiral galaxy Malin~1, and the Milky Way are used to
constrain colours at extremes of luminosity and surface brightness.
Because the colour of the old population is related to both luminosity
and surface brightness, the fraction of light from the old component
(Equation~\ref{eqn_fold}) must be determined through an iterative
process.
Based upon what is observed for old stars in elliptical galaxies,
the stellar mass--to--light ratio of the old stars is assumed to vary
as a power law in luminosity of the old component and is given by
\begin{equation}
\left( \frac{M_{\ast}}{L_B} \right)_{\rm old} =
\left( \frac{M_{\ast}}{L_B} \right)_{\rm old\,MW\,dsk}
\left( \frac{L_{B,{\rm old}}}{L_{B,{\rm old\,MW\,dsk}}} \right)
^{\gamma_d},
\label{eqn_mstarlb_old}
\end{equation}
where $(\frac{M_{\ast}}{L_B})_{\rm old\,MW\,dsk}$ and
$L_{B,{\rm old\,MW\,dsk}}$ are, respectively, the mass--to--light
ratio and luminosity in $B$ of the old component of the Milky Way
disk.
The mass--to--light ratio for the old component of the disk of the
Milky Way is obtained from the dispersion in vertical motions of
disk matter in proximity to the solar neighbourhood.
The power law exponent is judged from the relationship between the
stellar mass--to--light ratio and luminosity observed for the old
disks of spirals and dwarfs in the Virgo Cluster and is
found to be 
\begin{equation}
\gamma_d = 0.175
\label{eqn_gammad}
\end{equation}
(McCall 2000, private communication).
The zero point in Equation~(\ref{eqn_mstarlb_old}) is set by the
mass--to--light ratio for the old disk of the Milky Way, as judged
from the stellar kinematics perpendicular to the Galactic plane
\citep{gould90}. 
The derived stellar mass--to--light ratio and the luminosity of the
old Milky Way disk, respectively, are given by 
\begin{equation}
( M_{\ast}/L_B )_{\rm old\,MW\,dsk} = 3.11\;M_{\odot}/L_{B,{\odot}}
\label{eqn_mstarlb_oldmwdsk}
\end{equation}
and
\begin{equation}
L_{B,{\rm old\,MW\,dsk}} = 1.0 \times 10^{10}\;L_{B,{\odot}}
\label{eqn_lb_oldmw}
\end{equation}
(McCall 2000, private communication; see also \citealp{lee01}).
Combining Equations~(\ref{eqn_mstartotal}) through (\ref{eqn_lb_oldmw}),
the total mass of stars is
\begin{eqnarray}
M_{\ast} 
& = & M_{{\ast},{\rm yng}} + M_{{\ast},{\rm old}} \nonumber \\
& = & 0.153 \, (1 - f_{\rm old}) L_B \;+\;
3.11 \, \frac{ \left( f_{\rm old} L_B \right)^{1.175}}
                { \left( 1.0 \times 10^{10} \right)^{0.175} }.
\label{eqn_mstarfinal}
\end{eqnarray}

\begin{figure}
\epsscale{0.5}
\plotone{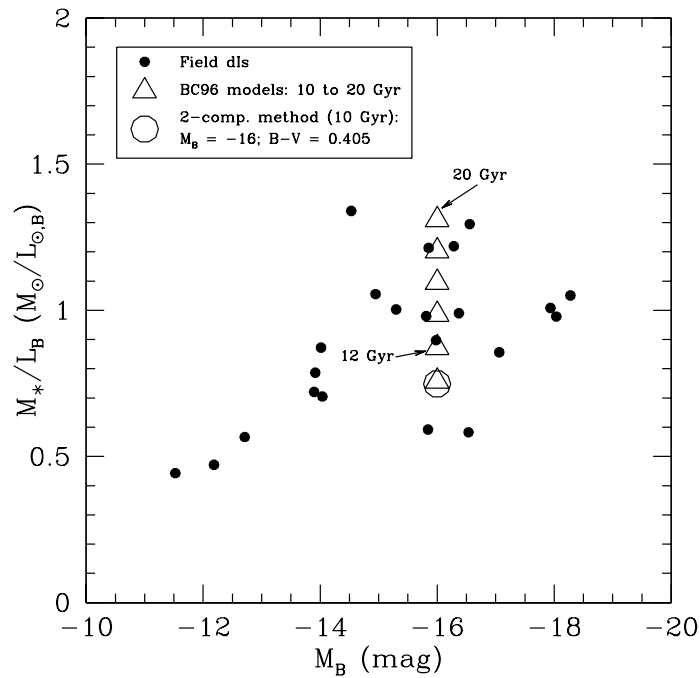}
\caption{
Stellar mass--to--light ratio versus absolute magnitude in $B$ 
for field dIs.
Filled circles mark the field dIs. 
For ages between 10 and 20 Gyr in increments of 2~Gyr, open triangles
show the locations of a fiducial $M_B = -16$ dwarf galaxy, whose
properties are determined from the evolutionary synthesis code of
\citet[BC96]{bc96}.
The open circle shows the position of a 10~Gyr old dI with $M_B = -16$
and $B-V = 0.405$ for which values of $M_{\ast}$ are obtained from the
two--component method.
}
\label{fig_mstarlb_mb}
\end{figure}

Figure~\ref{fig_mstarlb_mb} shows a plot of the stellar
mass--to--light ratio in $B$ versus absolute magnitude in $B$ for
field dIs. 
Two--component stellar mass--to--light ratios in $B$ for the sample of
field dIs range between 0.4 and 1.3, with most clustered around unity.
There is no strong evidence for a correlation of $M_{\ast}/L_B$ with
$M_B$. 

Stellar mass--to--light ratios computed with the two--component
method agree with predictions of full--scale population syntheses
\citep{bc93,bc96} for systems between 10 and 20 Gyr old which have
formed stars at a constant rate (specifically, for a Salpeter stellar
initial mass function with mass limits between 0.1~\msun\ and
125~\msun, and a metallicity equal to one--fifth solar). 
Assigning $M_B = -16$ to a dwarf, which is typical of the range
for this sample of dIs, the Bruzual \& Charlot models produce a dI
with $B-V = 0.405$ and $M_{\ast}/L_B = 0.757$ at an age of 10~Gyr.
For a fiducial dI with $M_B = -16$ and $B-V = 0.405$, the
two--component method yields $M_{\ast}/L_B = 0.748$, which is in
good agreement.

The gas fraction is given by
\begin{equation}
\mu = \frac{M_{\rm gas}}{M_{\rm bary}} 
    = \frac{M_{\rm gas}}{M_{\rm gas} + M_{\ast}}, 
\label{eqn_mu_defn}
\end{equation}
where the total mass in baryons, $M_{\rm bary}$, is taken to be the
mass of gas and stars.
The gas fraction is a distance--independent quantity, because the gas
mass and stellar mass are both derived from electromagnetic fluxes. 
Derived gas masses, stellar masses, stellar mass--to--light
ratios, and gas fractions for the sample of field dIs are listed in
Table~\ref{table_alldi_starsgas}.

\subsection{Oxygen Abundance versus Gas Fraction}
\label{sec_oxymu}

The ``closed box'' model serves as a useful guide to understanding the
chemical evolution of galaxies
\citep{ss72,pp75,at76,tinsley80,pe81,edmunds90,koppen93,pagel_book}.
Rate equations lead to the following relation between the fraction of
the gas mass in the form of a primary product of nucleosynthesis, $Z$
(i.e., the metal abundance), and the fraction of baryonic mass in
gaseous form, $\mu$, for a system which evolves in isolation (i.e.,
neither gaining nor losing mass):
\begin{equation}
Z = y\, \ln (1/\mu).
\label{eqn_closedbox1}
\end{equation}
The constant $y$ is called the yield, which is the ratio of the mass
of newly formed metals to the mass of gas permanently locked into
stars.
Additional premises for the model include an invariant stellar initial
mass function 
(e.g., \citealp{massey95a,massey95b,holtzman97,larson99})
and instantaneous recycling, which is valid for oxygen 
produced mainly in massive stars and returned to the interstellar
medium when these short--lived stars explode as Type~II supernovae
(e.g., \citealp{burrows00}). 

The metal abundance is best expressed as an elemental abundance in
logarithmic form.
For oxygen, Equation~(\ref{eqn_closedbox1}) is rewritten as
\begin{equation}
\log Z_{\rm O} = \log y_{\rm O} + \log \, \ln (1/\mu),
\end{equation}
where $Z_{\rm O}$ is the fraction of the gas mass in the form of oxygen.
The oxygen abundance by number is then given by
%
%
\begin{eqnarray}              
\lefteqn{
12 + \log \left[ N({\rm O})/N({\rm H}) \right] = 
12 + \log({\rm O/H}) 
}
\nonumber \\
& = & 12 + \log (2.303 \, y_{\rm O}/11.728) + \log \, \log (1/\mu),
\label{eqn_closedbox2}
\end{eqnarray}
where the numerical factor 2.303 comes from the conversion from a
natural logarithm to a logarithm of base ten and the numerical factor
11.728 comes from the conversion from an oxygen abundance by mass to
an oxygen abundance by number, assuming that the fraction of gas in
the form of hydrogen is $X = 0.733$ (solar value). 
In a plot of oxygen abundance, 12$+$log(O/H), against inverse gas mass
fraction, conveyed by $\log \log (1/\mu)$, the closed box model
predicts a slope of unity.
The oxygen yield, $y_{\rm O}$, can be derived from the intercept of the
plot.

\begin{figure}
\epsscale{0.7}
\plotone{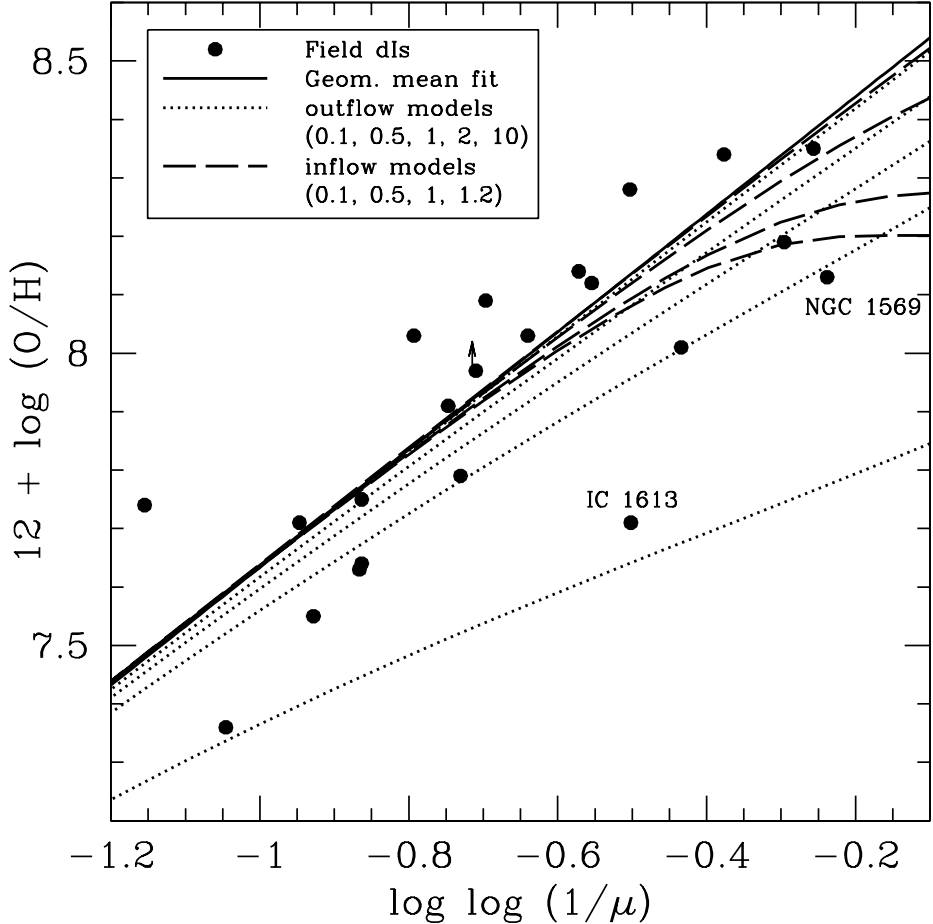}
\caption{
Oxygen abundance versus fraction of baryons in the form of gas 
for field dIs. 
The abundance increases upwards and gas fraction decreases
to the right.
Filled circles mark the field dIs.
Data points representing the dwarfs IC~1613 and NGC~1569 are labelled.
The upward arrow indicates a lower limit to the oxygen abundance for 
NGC~1560.
The solid line is a geometric mean fit to the field dIs
(Equation~\ref{eqn_zmu_fit}) consistent with the closed box model.
Dotted and dashed curves indicate different outflow and inflow models,
respectively.
The flow rate, which is taken to be proportional to the star formation
rate, increases from the top curve to the bottom curve in each set
of inflow and outflow models.
The error bars indicate typical uncertainties of 0.1~dex in oxygen
abundance and 0.1~dex in log log~(1/$\mu$). 
The latter uncertainty is derived from an estimated 0.05~mag
uncertainty in $B-V$, which affects the derivation of $M_{\ast}$,
and an estimated 20\% uncertainty in \hi\ gas mass 
\citep{hr86}.
}
\label{fig_zmu_field}
\end{figure}

A plot of oxygen abundance against inverse gas mass fraction for field
dIs is shown in Figure~\ref{fig_zmu_field}, where an excellent
correlation is seen.
A fit to all field dIs gives 
\begin{equation}
12 + \log{\rm (O/H)} = 
        (8.63 \pm 0.40) + (1.02 \pm 0.21)\,\log \,\log (1/\mu).
\end{equation}
The rms in log(O/H) is $\sigma = 0.183$~dex.

IC~1613 appears to have an anomalously small gas mass
fraction (log log (1/$\mu$) $\simeq -0.5$) and/or a small abundance
(12$+$log(O/H) $\simeq 7.7$).
The \hi\ mass is unlikely to be problematic, because the most
recent 21--cm flux measurement obtained by \cite{hoffman96} has
been employed. 
The measurement of the oxygen abundance may be problematic, however, 
because the most recent spectroscopic observations are over
two decades old \citep{talent80}. 
Another problem may be the reported $B-V$ colour.
If it appears too red for its luminosity, the mass judged for the
underlying old stellar population would be an overestimate, and, in
turn, the gas fraction would be an underestimate.

Excluding IC~1613, the fit is
\begin{equation}
12 + \log ({\rm O/H}) = 
        (8.64 \pm 0.40) + (1.01 \pm 0.17)\,\log \,\log (1/\mu).
\label{eqn_zmu_fit}
\end{equation}
The rms in log(O/H) is reduced to $\sigma = 0.162$~dex, which
is lower than the rms in log(O/H) determined for the fit to the
oxygen abundance versus luminosity relationship
(Figure~\ref{fig_zmb_field}; Equation~\ref{eqn_zmb_new}).
Equation~(\ref{eqn_zmu_fit}) is taken as the ``best fit'' 
and is shown as a solid line in Figure~\ref{fig_zmu_field}. 
The fitted slope is consistent with the prediction of the
closed box model in Equation~(\ref{eqn_closedbox2}).

Comparing the intercept in Equation~(\ref{eqn_zmu_fit})
with Equation~(\ref{eqn_closedbox2}), the effective oxygen yield 
by mass, $y_{\rm O}$, is 
\begin{equation}
y_{\rm O} = 2.22 \times 10^{-3}.
%
\label{eqn_oxyyield}
\end{equation}
This result is consistent with that obtained from a fit to a mix
of gas--rich dwarf galaxies compiled by \cite{pagel86}. 
If $Z_{\odot,{\rm O}} = 0.02 \times 0.45 = 0.009$, the present
result for the oxygen yield is one--quarter of the value for the solar 
neighbourhood\footnote{ 
The solar value of the oxygen abundance of 12$+$log(O/H) = 8.87
is adopted for the present work \citep{zsolar}.
However, recent results (e.g., \citealp{zsolarnew}) have shown that
the solar value may in fact be smaller by $\approx$~0.2~dex.
}.

We constucted additional models of chemical evolution by extending
the closed box model by accounting for flows; see recent work by,
e.g., \cite{edmunds90,pagel_book,rms98,rms01}.
A full description of the latter will be given by \cite{mrs03}.
In short, the flow rate is a multiple of the star formation rate
(i.e., linear flows, constant with time; see e.g.,
\citealp{mc83,edmunds90,koppen93}).
In the analytical formulation, the explicit time dependence
cancels out.
The metallicity of the flowing gas is either zero for inflow, or
equal to the metallicity of the interstellar medium for outflow.  
The adopted yield is the same derived above in
Equation~(\ref{eqn_oxyyield}).

The predicted relations from models with gas flows are shown in
Figure~\ref{fig_zmu_field}. 
Outflow and inflow models are indicated as dotted and dashed curves,
respectively.
In each set of models, the flow rate increases from the top curve to
the bottom curve.
For outflow models, the slopes of the model curves decrease as
the flow rates increase.
For inflow rates higher than the star formation rate, the shape of 
the curve deviates from a straight line and turns over at low gas
fractions. 
Nonzero flow rates exceeding the star formation rate are allowed,
but not forced.
The simplest explanation consistent with the data is that the flow
rates are zero (i.e., dIs as closed systems), although the
observations admit the possibility that there were moderate flows.
It would be interesting to augment the number of metal-poor dwarf
irregular galaxies to help ``anchor'' the slope at high gas
fractions and to augment the number of dwarf galaxies with
\othreea\ abundances at lower gas fractions to see whether
the trend ``turns over'' towards nonzero flows.

The value of the metallicity-gas fraction diagram in studying chemical
evolution is that statements about chemical evolution, especially the
importance of gas flows, can be made without knowledge of the detailed
history of star formation.
On its own, the metallicity-luminosity diagram says nothing about
flows.
It is simply evidence of a relationship between metallicity and a
parameter that depends upon luminosity.
For closed box evolution, that parameter could be (a) the average star
formation rate, if all dIs have been forming stars for the same length
of time, 
or (b) the star formation time scale, if all dIs have been
forming stars at the same average rate, 
or (c) some combination of the two for more complicated star 
formation scenarios.  
Recent studies (e.g., \citealp{ght84,vanzee01}) have shown that stars
in dwarf irregulars have formed steadily but slowly.   
Since the metallicity-gas fraction diagram suggests that field dwarfs
have not suffered from extreme flows of gas, then one would conclude
that the metallicity-luminosity relation is a consequence of
variations in the time scale for star formation. 


Values of $\mu$ depend on $M_{\ast}$, which in turn comes from the
two--component algorithm of population synthesis.
It is worthwhile examining the ingredients of the algorithm to
determine the robustness of the conclusion that field dIs have evolved
as closed systems. 
Four variants are considered: 
(1) the mass--to--light ratio for old stars is fixed at the value
for the old disk of the the Milky Way
(see Equation~\ref{eqn_mstarlb_oldmwdsk}),
(2) the mass--to--light ratio for old stars is fixed at the mean
value (1.50) for all dIs in the sample,
(3) the mass--to--light ratio for old stars varies with the luminosity
of the old component as a power law, and the $B-V$ colour for old
stars is fixed at the mean value (0.74) for all dIs in the sample, 
and 
(4) the mass--to-light ratio for {\em all\/} stars is fixed
at unity, motivated by the \cite{bc96} models and the absence of any
trend in $B-V$ colour with $B$ luminosity \citep{lee01}.
In each case, $M_{\ast}$ and $\mu$ were computed, and a best fit 
to log(O/H) vs. log~log~(1/$\mu$) was determined using the geometric
mean functional relationship. 


Table~\ref{table_const_mstarlb} lists the effects of the presumptions
on the relationship described by Equation~(\ref{eqn_closedbox2}).
The exclusion of the data point corresponding to IC~1613 
from all models does not significantly affect the results.
All fits have slopes consistent with unity.
The smallest dispersion in the relationship between the oxygen
abundance and the gas fraction is obtained when the mass--to--light
ratio in the old disk component has a power--law dependence on the
luminosity of the old disk, as employed in the two--component
algorithm.  
Hence, the conclusion that field dIs have evolved as closed systems
does not depend critically upon the details of the population
synthesis.

\section{Conclusions}		
\label{sec_concl}		

The evolution of dwarf irregular galaxies in the field must be
understood before it is possible to study how the evolution of dIs in
clusters of galaxies has been affected by the denser environment.
A suitable control group was constructed from a sample of nearby
dIs with \othreea\ measurements and well--measured distances. 
Distance determinations and oxygen abundance measurements were
updated with recently published values.
Oxygen abundances for five field dIs were updated using spectroscopy
acquired at Steward Observatory and at the Observatorio Astron\'omico
Nacional in M\'exico. 

Two diagrams were examined as diagnostics for the evolution of dIs.
The metallicity--luminosity relationship has been considered as a
proxy for the metallicity--mass relationship, at least where stellar
mass is concerned.
The updated relationship between oxygen abundance and luminosity
for the present sample of field dIs is similar to that found by
\cite{skh89} and \cite{rm95}.
The relationship between metallicity and the gas fraction is a gauge of
the progress by which gas is being converted into stars and metals.
The gas fraction was judged using the masses of luminous components in
the form of stars and gas, which are constituents of the nucleogenetic
pool. 
This is preferable to the past use of dynamical masses that may 
be dominated by non--baryonic dark matter, which does not participate
in nucleosynthesis. 
A strong correlation between oxygen abundance and the gas fraction,
defined as the ratio of gas mass to the total baryonic mass, was found
for field dIs. 
The data are consistent with zero gas flows (i.e., evolution as
isolated systems), although the observations do admit the possibility
of modest flows.
The oxygen yield is about one--quarter of the value found in the solar
neighbourhood.

It is now possible to use the diagnostic diagrams to study the
evolution of dwarfs in clusters of galaxies by looking for an offset
between a sample of cluster dwarfs and a control sample of dwarfs in
the field. 
The present results are applied to a study of dwarf irregulars in the
Virgo Cluster, which is the subject of Paper II.

\acknowledgments		

The work presented here was part of the dissertation completed by HL
at York University in Toronto, Canada.
HL is grateful to Marshall McCall for guidance, supervision, and
financial support.  
HL also thanks Dan Zucker, Michael Richer, and the anonymous referee
for constructive comments which improved the presentation of this
paper.
MLM acknowledges the Natural Sciences and Engineering Research Council 
of Canada for its continuing support.
MLM, RLK, and CCS thank the staff at the Observatorio Astron\'omico
Nacional at San Pedro M\'artir and the Steward Observatory for their
help with observations.
Some data were accessed as Guest User, Canadian Astronomy Data Center,
which is operated by the Dominion Astrophysical Observatory for the 
National Research Council of Canada's Herzberg Institute of
Astrophysics.
This research has made use of the NASA/IPAC Extragalactic Database
(NED), which is operated by the Jet Propulsion Laboratory,
California Institute of Technology, under contract with the
National Aeronautics and Space Administration.

\clearpage	


\clearpage	

\begin{deluxetable}{cccccc}
\renewcommand{\arraystretch}{1.}
\tablecolumns{6}
\tabletypesize{\scriptsize} 
\tablewidth{0pt}
\tablecaption{
The sample of dwarf irregular galaxies in the field.
\label{table_fieldsample}}
\tablehead{
\colhead{Name} & \colhead{$M_B$} & \colhead{$(m-M)^0$} & 
\colhead{$(B-V)_T^0$} & \colhead{$F_{21}$} & \colhead{12$+$log(O/H)} \\
& \colhead{(mag)} & \colhead{(mag)} & \colhead{(mag)} & 
\colhead{(Jy km s$^{-1}$)} & \colhead{(dex)} \\
\colhead{(1)} & \colhead{(2)} & \colhead{(3)} & 
\colhead{(4)} & \colhead{(5)} & \colhead{(6)}
}
\startdata
\objectname[]{DDO 187}      & $-$12.72 & 27.10   & 0.38  & 14.2 &
        $7.69 \pm 0.09$ \\        
\objectname[]{GR 8}         & $-$12.19 & 26.83   & 0.31  & 8.6 &
        $7.63 \pm 0.14$ \\        
\objectname[]{Holmberg II}  & $-$15.98 & 27.50   & 0.42  & 359.7 & 
        $7.71 \pm 0.13$ \\        
\objectname[]{IC 10}$\,$\tablenotemark{a} & 
               $-$15.85 & 24.47  & 0.69  & 950.4 &
        $8.19 \pm 0.14$ \\        
\objectname[]{IC 1613}      & $-$14.53 & 24.38  & 0.67  & 698 &
        $7.71 \pm 0.15$ \\        
\objectname[]{IC 2574}      & $-$17.06 & 27.86   & 0.35  & 442.5 &
        $8.09 \pm 0.07$ \\ 
\objectname[]{IC 4662}      & $-$15.84 & 27.32  & 0.35  & 125.2 &
        $8.09 \pm 0.04$ \\        
\objectname[]{Leo A}        & $-$11.53 & 24.45   & 0.32  & 68.4 &
        $7.36 \pm 0.06$ \\        
\objectname[]{LMC}          & 
               $-$17.94 & 18.58   & 0.45  & $1.044 \times 10^6$ & 
        $8.35 \pm 0.06$ \\        
\objectname[]{NGC 55}       & $-$18.28 & 25.94  & 0.46  & 2679 & 
        $8.34 \pm 0.10$ \\        
\objectname[]{NGC 1560}     & $-$16.37 & 27.07  & 0.49  & 443 &
        $> 7.97$ \\               
\objectname[]{NGC 1569}$\,$\tablenotemark{b} & 
               $-$16.54 & 26.39  & 0.35  & 116 &
        $8.19 \pm 0.06$ \\ 
\objectname[]{NGC 2366}     & $-$16.28 & 27.76   & 0.54  & 297.0 &
        $7.91 \pm 0.08$ \\        
\objectname[]{NGC 3109}     & $-$15.30 & 25.72   & 0.49  & 1876.7 &
        $7.74 \pm 0.33$ \\        
\objectname[]{NGC 4214}     & $-$18.04 & 28.25  & 0.46  & 368.8 &
        $8.24 \pm 0.12$ \\ 
\objectname[]{NGC 5408}     & $-$15.81 & 27.76  & 0.49   & 59.20 &
        $8.01 \pm 0.02$ \\        %
\objectname[]{NGC 6822}     & $-$14.95 & 23.46  & 0.59  & 2367.1 &
        $8.25 \pm 0.11$ \\ 
\objectname[]{Sextans A}    & $-$14.04 & 25.84   & 0.38  & 208.8 &
        $7.55 \pm 0.11$ \\        
\objectname[]{Sextans B}    & $-$14.02 & 25.67   & 0.51  & 102.4 &
        $8.12 \pm 0.12$ \\        
\objectname[]{SMC}          & 
               $-$16.56 & 19.06   & 0.59  & $8.99 \times 10^5$ & 
        $8.03 \pm 0.10$ \\        
\objectname[]{UGC 6456}     & $-$13.90 & 28.31  & 0.38   & 16.20 &
        $7.64 \pm 0.13$ \\        
\objectname[]{WLM}          & $-$13.92 & 24.86  & 0.42  & 299.8 &
        $7.78 \pm 0.16$ \\ 
\enddata
\tablenotetext{a}{
Its optical appearance shares similarities with blue compact dwarfs
\citep{richer01}.
}
\tablenotetext{b}{
This is a post--starburst dwarf galaxy \citep{heckman95,greggio98}. 
}
\tablecomments{
Column (1): Galaxy name in alphabetical order
Column (2): Absolute magnitude in $B$.
Column (3): Extinction--corrected distance modulus.
Column (4): Total dereddened $B-V$ colour.
Column (5): Integrated flux density at the 21--cm \hi\ line.
Column (6): Logarithm of the oxygen abundance by number.
}
\end{deluxetable}


\begin{deluxetable}{cccc}
\renewcommand{\arraystretch}{1.}
\tablecolumns{4}
\tabletypesize{\scriptsize} 
\tablewidth{0pt}
\tablecaption{
Revisions to distances for the sample of dIs in the field.
\label{table_distupdate}}
\tablehead{
\colhead{Galaxy} & \colhead{New/Update} & 
\colhead{Method$\,$\tablenotemark{a}} & \colhead{References} \\
\colhead{(1)} & \colhead{(2)} & \colhead{(3)} & \colhead{(4)}
}
\startdata                  
DDO 187 & New & Ceph, TRGB & 1, 2 \\               
GR 8 & Update & Ceph, TRGB & 3, 4 \\
Holmberg II & Update & Ceph & 5 \\
IC 10 & Update & Ceph, TRGB & 6, 7, 8 \\
IC 1613 & Update & TRGB & 9 \\
Leo A & Update$\,$\tablenotemark{b} & Ceph, TRGB & 10, 11 \\  
LMC & Update & SN87, TRGB & 12, 13 \\     
NGC 2366 & Update & Ceph & 14 \\          
NGC 3109 & Update & Ceph, TRGB & 15, 16 \\  
NGC 6822 & Update & Ceph, TRGB & 17 \\     
UGC 6456 & Update & TRGB & 18, 19 \\       
Sextans A & Update & Ceph, TRGB & 20 \\   
Sextans B & Update & Ceph, TRGB & 21 \\   
WLM & Update & TRGB & 22 \\ 
\enddata
\tablenotetext{a}{
Distance method: 
``Ceph'' denotes Cepheid variables,
``SN87'' denotes the expanding ring of the supernova remnant SN1987A, and 
``TRGB'' denotes the tip of the red giant branch.
}
\tablenotetext{b}{
For Leo A, Tolstoy et al. (1998) dispute the claim of detected
Cepheids by Hoessel et al. (1994).
We adopt here the TRGB distance obtained by Tolstoy et al. (1998).
}
\tablecomments{
Column (1) : Galaxy name in alphabetical order.
Column (2) : New addition or update. 
Column (3) : Method by which distances are obtained.
Column (4) : References.
}
\tablerefs{
(1) Hoessel et al. (1998b);	
(2) Aparicio et al. (2000);
(3) Dohm--Palmer et al. (1998);
(4) Tolstoy et al. (1995a);	
(5) Hoessel et al. (1998a); 	
(6) Saha et al. (1996);
(7) Sakai et al. (1999);
(8) Wilson et al. (1996);
(9) Cole et al. (1999);
(10) Hoessel et al. (1994);
(11) Tolstoy et al. (1998);
(12) Panagia (1999);
(13) Sakai et al. (2000);
(14) Tolstoy et al. (1995b);	
(15) Minniti et al. (1999);
(16) Musella et al. (1997);
(17) Gallart et al. (1996);
(18) Lynds et al. (1998);
(19) Schulte--Ladbeck et al. (1998);
(20) Sakai et al. (1996);
(21) Sakai et al. (1997);
(22) Minniti \& Zijlstra (1997).
}
\end{deluxetable}


\begin{deluxetable}{ccc}
\renewcommand{\arraystretch}{1.}
\tablecolumns{3}
\tabletypesize{\footnotesize} 
\tablewidth{0pt}
\tablecaption{
Revisions to oxygen abundances for the sample of dIs in the field.
\label{table_specupdate}}
\tablehead{
\colhead{Galaxy} & \colhead{New/Update} & \colhead{References} \\
\colhead{(1)} & \colhead{(2)} & \colhead{(3)}
}
\startdata             
DDO 187 & New & 1 \\             
Holmberg II & Update & 2 \\       
IC 10 & Update & 2 \\             
IC 1613 & Update & 3 \\           
IC 2574 & Update & 4 \\           
NGC 1560 & Update & 2 \\          
NGC 1569 & Update & 5 \\          
NGC 2366 & Update & 6, 7 \\       
NGC 3109 & Update & 2 \\          
NGC 4214 & Update & 8 \\          
NGC 6822 & Update & 9 \\          
UGC 6456 & Update & 2, 7, 10 \\   
WLM & Update & 11 \\              
\enddata
\tablecomments{
Column (1) : Galaxy name in alphabetical order.
Column (2) : New addition or update. 
Column (3) : References.
}
\tablerefs{
(1) \cite{vanzee97};               
(2) present work;
(3) \cite{kb95};                   
(4) \cite{mh96};                   
(5) \cite{ks97};                   
(6) \cite{gd94};                   
(7) \cite{itl97};                  
(8) \cite{ks96};                   
(9) \cite{miller96};               
(10) \cite{hh99};                  
(11) \cite{hm95}.                  
}
\end{deluxetable}


\begin{table}
\footnotesize
\begin{center}
\renewcommand{\arraystretch}{1.1}
\caption{Instrument configurations at Steward and Observatorio
Astron\'omico Nacional (OAN). 
\vspace*{3mm}
\label{table_stewardspm}
}
\begin{tabular}{cccc}
\tableline \tableline
Property & Steward (blue) & Steward (red) & 
OAN (red) \\ \tableline
\multicolumn{4}{c}{{\sf Telescope}} \\ \tableline
\multicolumn{1}{c}{Aperture} & 
\multicolumn{2}{c}{2.3 metre Bok} &
\multicolumn{1}{c}{2.1 metre} \\ 
\multicolumn{1}{c}{Focus} & 
\multicolumn{2}{c}{f/9.0} & 
\multicolumn{1}{c}{f/7.5} \\ 
\multicolumn{1}{c}{Spectrograph} & 
\multicolumn{2}{c}{Boller \& Chivens} & 
\multicolumn{1}{c}{Boller \& Chivens} \\ \tableline
\multicolumn{4}{c}{{\sf CCD}} \\ \tableline
\multicolumn{1}{c}{Type} & 
\multicolumn{2}{c}{Texas Instruments} & 
\multicolumn{1}{c}{Tektronix} \\ 
\multicolumn{1}{c}{Total area} & 
\multicolumn{2}{c}{800 pix $\times$ 800 pix} & 
\multicolumn{1}{c}{1024 pix $\times$ 1024 pix} \\ 
\multicolumn{1}{c}{Usable area} & 
\multicolumn{2}{c}{800 pix $\times$ 350 pix} & 
\multicolumn{1}{c}{1024 pix $\times$ 400 pix} \\
\multicolumn{1}{c}{Pixel size} & 
\multicolumn{2}{c}{15 $\mu$m} & 
\multicolumn{1}{c}{24 $\mu$m} \\
\multicolumn{1}{c}{Image scale} & 
\multicolumn{2}{c}{0.8 arcsec pix$^{-1}$} & 
\multicolumn{1}{c}{0.92 arcsec pix$^{-1}$} \\
\multicolumn{1}{c}{Gain} & 
\multicolumn{2}{c}{2.8 $e^-$ ADU$^{-1}$} & 
\multicolumn{1}{c}{1.22 $e^-$ ADU$^{-1}$} \\ 
\multicolumn{1}{c}{Read--noise (rms)} & 
\multicolumn{2}{c}{7.4 $e^-$} & 
\multicolumn{1}{c}{3 $e^-$} \\ \tableline
\multicolumn{4}{c}{{\sf Grating, Long slit}} \\ \tableline
Groove density & 
600 lines mm$^{-1}$ & 600 lines mm$^{-1}$ & 300 lines mm$^{-1}$ \\
Blaze $\lambda$ (1st order) & 3568 \AA & 6690 \AA & 5000 \AA \\ 
Dispersion & 
1.89 \AA\ pix$^{-1}$ & 3.72 \AA\ pix$^{-1}$ & 3.96 \AA\ pix$^{-1}$ \\ 
Effective $\lambda$ range &
3650--5100 \AA & 4200--7200 \AA & 3450--7450 \AA \\
Slit width & 2.5 arcsec & 4.5 arcsec & 2.2 arcsec \\ 
\tableline
\end{tabular}
\end{center}
\end{table}


\begin{deluxetable}{ccccccc}
\renewcommand{\arraystretch}{1.}
\tablecolumns{7}
\tabletypesize{\footnotesize} 
\tablewidth{0pt}
\tablecaption{
Log of Observations.
\label{table_obslog}}
\tablehead{
\colhead{Galaxy} & \colhead{Obs.} & \colhead{Date} &
\colhead{$N_{\rm exp}$} & \colhead{$t_{\rm total}$} & 
\colhead{\protect \othreea} & \colhead{$\delta$} \\
& & \colhead{(UT)} & & \colhead{(s)} & & \colhead{(mag)} \\
\colhead{(1)} & \colhead{(2)} & \colhead{(3)} & \colhead{(4)} &
\colhead{(5)} & \colhead{(6)} & \colhead{(7)}
}
\startdata
Holmberg II & Steward & 1992 Mar 23 & 
	1$\times$300 $+$ 3$\times$1200 &
        3900 & detected & 3.7 \% \\
IC 10 & Steward & 1991 Oct 15 & 
	1$\times$60 $+$ 3$\times$1800 &
        5460 & detected & 2.2 \% \\
IC 10 & OAN$\,$\tablenotemark{a} & 1994 Nov 30 & 1$\times$1200 &
        1200 & \nodata & 3.0 \% \\
NGC 1560 & Steward & 1991 Oct 15 & 3$\times$1800 &
        5400 & upper limits & 2.2 \% \\
NGC 1560 ``1'' (NE)$\,$\tablenotemark{b} & Steward$\,$\tablenotemark{a} & 
	1992 Jan 27 & 1$\times$1200 & 1200 & \nodata & 2.2 \% \\
NGC 1560 ``2'' (SW)$\,$\tablenotemark{c} & Steward$\,$\tablenotemark{a} & 
	1992 Jan 27 & 1$\times$900 & 900 & \nodata & 2.2 \% \\
NGC 3109 & Steward & 1992 Mar 23 & 
	1$\times$300 $+$ 4$\times$1200 &
        5100 & detected & 3.7 \% \\
UGC 6456 & Steward & 1992 Mar 23 & 
	1$\times$300 $+$ 4$\times$900 &
        3900 & detected & 3.7 \% \\
\enddata
\tablenotetext{a}{Designed to measure ratio of H$\alpha$/H$\beta$
fluxes.}
\tablenotetext{b}{One exposure for slit placement to the
northeast of the main body of the galaxy.}
\tablenotetext{c}{One exposure for slit placement to the
southwest of the main body of the galaxy.}
\tablecomments{
Column (1): Galaxy name.
Column (2): Observatory.
Column (3): Date.
Column (4): Number of exposures obtained and the length of each
exposure in seconds. 
Column (5): Total exposure time.
Column (6): \protect \othreea\ detection.
Column (7): Relative root--mean--square error in the sensitivity
function obtained from observations of standard stars.
}
\end{deluxetable}


\begin{deluxetable}{cccc}
\renewcommand{\arraystretch}{1.}
\tablecolumns{4}
\tabletypesize{\footnotesize} 
\tablewidth{0pt}
\tablecaption{
Identification of H II regions in Holmberg II (Ho II).
\label{table_holm2_h2rs}}
\tablehead{
\colhead{H II Region} & \colhead{\protect \othreea} & 
\colhead{HSK94} & \colhead{HG85} \\
\colhead{(1)} & \colhead{(2)} & \colhead{(3)} & \colhead{(4)} 
}
\startdata
Ho II--1 & upper limit & 67\phn & 6 \\ 
Ho II--2$\,$\tablenotemark{a} & upper limit & 67\phn & 6 \\
Ho II--3 & upper limit & 67\phn & 6 \\ 
Ho II--4 & upper limit & 69? & \nodata \\ 
Ho II--5 & detected & 70\phn & 5 \\ 
Ho II--6 & upper limit & 73\phn & 4 \\ 
Ho II--7$\,$\tablenotemark{b} & detected & 73\phn & 4 \\ 
Ho II--8 & detected & 73\phn & 4 \\ 
Ho II--9 & detected & 71\phn & 3 \\ 
\enddata
\tablenotetext{a}{
The \hii\ region Ho~II--2 encompasses two separate \hii\ regions 
1 and 3.  
Subsequent line intensities and derived quantities for Ho~II--2 are
not independent of the two latter \hii\ regions.
}
\tablenotetext{b}{
The \hii\ region Ho~II--7 encompasses two separate \hii\ 
regions 6 and 8. 
Subsequent line intensities and derived quantities for Ho~II--7 are
not independent of the two latter \hii\ regions.
}
\tablecomments{
Column (1): \hii\ region.  
The \hii\ region number increases along the slit towards the north.
Column (2): \othreea\ detection.
Columns (3) and (4): corresponding \hii\ regions identified by
\citet[Figure 2, HSK94]{hsk94} and 
\citet[Figure 1 for DDO~50, HG85]{hg85}, respectively.
}
\end{deluxetable}


\begin{deluxetable}{cccc}
\renewcommand{\arraystretch}{1.}
\tablecolumns{4}
\tabletypesize{\footnotesize} 
\tablewidth{0pt}
\tablecaption{
Identification of H II regions in IC 10.
\label{table_ic10_h2rs}}
\tablehead{
\colhead{H II Region} & \colhead{\protect \othreea} & 
\colhead{LPRSTP} & \colhead{HL90} \\
\colhead{(1)} & \colhead{(2)} & \colhead{(3)} & \colhead{(4)} 
}
\startdata
IC 10--1$\,$\tablenotemark{a} & detected & 1 & 111c,e \\ 
IC 10--2 & detected & \nodata & 111c \\ 
IC 10--3 & detected & \nodata & 111e \\ 
IC 10--4$\,$\tablenotemark{b} & upper limit & \nodata & 111b \\ 
IC 10--5 & upper limit & \nodata & 106a \\ 
IC 10--6$\,$\tablenotemark{c} & \nodata & \nodata & 
	106 south, diffuse \\ 
\enddata
\tablenotetext{a}{
The \hii\ region IC~10--1 encompasses two separate \hii\ regions 
2 and 3.  
Subsequent line intensities and derived quantities for IC~10--1 are
not independent of the two latter \hii\ regions.
}
\tablenotetext{b}{
Light from this \hii\ region may include a contribution from the
neighbouring \hii\ region 111a \citep{hl90}.
}
\tablenotetext{c}{
The observed spectrum was not of sufficient quality to be
included in the analysis.
}
\tablecomments{
Column (1): \hii\ region.  
The \hii\ region number increases along the slit towards the south.
Column (2): \othreea\ detection.
Columns (3) and (4): corresponding \hii\ regions identified by
\citet[Figure 1, LPRSTP]{lequeux79} and
\citet[Figure 2, HL90]{hl90}, respectively.
}
\end{deluxetable}


\begin{deluxetable}{ccc}
\renewcommand{\arraystretch}{1.}
\tablecolumns{3}
\tabletypesize{\footnotesize} 
\tablewidth{0pt}
\tablecaption{
Identification of H II regions in NGC 1560.
\label{table_n1560_h2rs}}
\tablehead{
\colhead{H II Region} & \colhead{\protect \othreea} & 
\colhead{Notes} \\
\colhead{(1)} & \colhead{(2)} & \colhead{(3)} 
}
\startdata
NGC 1560--1 & upper limit & MLM \#1$\,$\tablenotemark{a} \\ 
NGC 1560--2 & upper limit & bright \\ 
NGC 1560--3 & upper limit & faint \\ 
NGC 1560--4$\,$\tablenotemark{b} & upper limit & faint \\ 
NGC 1560--5 & upper limit & faint \\ 
NGC 1560--6 & not detected & very faint \\ 
NGC 1560--7 & not detected & very faint \\ 
\enddata
\tablenotetext{a}{
Identified as \hii\ region \#1 in the \halpha\ image by M. L. McCall.
}
\tablenotetext{b}{
The \hii\ region NGC~1560--4 encompasses two separate \hii\ regions
3 and 5.
Subsequent line intensities and derived quantities for NGC~1560--4 are
not independent of the two latter \hii\ regions.
}
\tablecomments{
Column (1): \hii\ region.
The \hii\ region number increases along the slit towards the northeast.
Column (2): \othreea\ detection.
Column (3): comments about the relative brightness observed
for each \hii\ region.
}
\end{deluxetable}


\begin{deluxetable}{cccc}
\renewcommand{\arraystretch}{1.}
\tablecolumns{4}
\tabletypesize{\small} 
\tablewidth{0pt}
\tablecaption{
Identification of H II regions in NGC 3109.
\label{table_n3109_h2rs}}
\tablehead{
\colhead{H II Region} & \colhead{\protect \othreea} & 
\colhead{Notes} & \colhead{RM92} \\
\colhead{(1)} & \colhead{(2)} & \colhead{(3)} & \colhead{(4)} 
}
\startdata
NGC 3109 -- 1 & no H$\gamma$ & very faint & \nodata \\
NGC 3109 -- 2 & no [O III]$\lambda$ 4959 & very faint & \nodata \\
NGC 3109 -- 3$\,$\tablenotemark{a} & detected & bright & 5 \\
NGC 3109 -- 4$\,$\tablenotemark{b} & upper limit & bright & 5 \\
\enddata
\tablenotetext{a}{
The extraction window was narrow (6~pixels) to isolate the signal
from the \hii\ region.
}
\tablenotetext{b}{
The extraction window was wide (20~pixels), which included the
spectrum of an early--type field star adjacent to the \hii\ region.
}
\tablecomments{
Column (1): \hii\ region.
The \hii\ region number increases along the slit towards the north.
Column (2): \othreea\ detection.
Column (3): notes about the \hii\ regions.
Column (4): corresponding \hii\ region identified by
\citet[Fig. 1, RM92]{rm92}.
}
\end{deluxetable}


\begin{deluxetable}{cccc}
\renewcommand{\arraystretch}{1.}
\tablecolumns{4}
\tabletypesize{\small} 
\tablewidth{0pt}
\tablecaption{
Identification of H II regions in UGC 6456.
\label{table_u6456_h2rs}}
\tablehead{
\colhead{H II Region} & \colhead{\protect \othreea} & 
\colhead{Lynds et al. 1998} & \colhead{Tully et al. 1981} \\
\colhead{(1)} & \colhead{(2)} & \colhead{(3)} & \colhead{(4)} 
}
\startdata
UGC 6456 -- 1 & detected & 5 & 2nd clump in slit \\ 
UGC 6456 -- 2 & detected & 1 & brightest clump in slit \\
\enddata
\tablecomments{
Column (1): \hii\ region.
The \hii\ region number increases along the slit towards the east.
Column (2): \othreea\ detection.
Columns (3) and (4): corresponding \hii\ regions identified by
\citet[Figure 2]{lynds98} and \citet[Figure 2]{tully81},
respectively.
}
\end{deluxetable}

\clearpage 

\begin{table}
\scriptsize
\begin{center}
\renewcommand{\arraystretch}{1.}
\caption{
Observed and corrected line ratios for Holmberg II.
\vspace*{3mm}
\label{table_holm2_data}
}
\begin{tabular}{ccccccc}
\tableline \tableline
& \multicolumn{2}{c}{Ho II--1} & 
\multicolumn{2}{c}{Ho II--2} &
\multicolumn{2}{c}{Ho II--3} \\
\cline{2-7}
\multicolumn{1}{c}{Identification (\AA)} &
\multicolumn{1}{c}{$F$} & \multicolumn{1}{c}{$I$} &
\multicolumn{1}{c}{$F$} & \multicolumn{1}{c}{$I$} &
\multicolumn{1}{c}{$F$} & \multicolumn{1}{c}{$I$} \\ 
\tableline
$[\rm{O\;II}]\;3727$ &
        $447 \pm 11$ & $443 \pm 48$ & 
        $259.3 \pm 4.7$ & $257 \pm 22$ & 
        $186.7 \pm 4.5$ & $185 \pm 16$ 
\\
$[\rm{Ne\;III}]\;3869$ &
        $48.8 \pm 4.8$ & $48.3 \pm 8.1$ &
        $34.3 \pm 2.6$ & $34.0 \pm 4.4$ &
        $29.5 \pm 2.6$ & $29.2 \pm 4.1$ 
\\ 
${\rm He\;I} + {\rm H8}\;3889$ &
        $34.1 \pm 4.6$ & $36.6 \pm 8.0$ &
        $26.4 \pm 2.5$ & $28.8 \pm 4.9$ &
        $25.1 \pm 2.6$ & $27.3 \pm 4.6$
\\ 
$[{\rm Ne\;III}] + {\rm H}\epsilon\;3970$ &
        $48.6 \pm 4.8$ & $50.4 \pm 9.0$ &
        $37.7 \pm 2.6$ & $39.1 \pm 4.6$ &
        $33.3 \pm 3.0$ & $34.5 \pm 5.2$ 
\\ 
${\rm H}\delta\;4101$ &
        $44.4 \pm 7.7$ & $46 \pm 12$ &
        $37.2 \pm 3.7$ & $38.9 \pm 6.2$ &
        $34.8 \pm 3.3$ & $36.5 \pm 5.7$
\\ 
${\rm H}\gamma\;4340$ &
        $61.2 \pm 4.4$ & $61.6 \pm 9.3$ &
        $58.4 \pm 2.4$ & $59.3 \pm 6.1$ &
        $57.1 \pm 2.4$ & $58.2 \pm 6.0$
\\ 
$[{\rm O\;III}]\;4363$ &
        $<$ 9.1 (2$\sigma$) & $<$ 6.8 (2$\sigma$) &
        $8.3 \pm 1.9$ & $8.2 \pm 2.3$ &
        $9.0 \pm 2.0$ & $8.9 \pm 2.4$ 
\\ 
${\rm H}\beta\;4861$ &
        $100.0 \pm 4.5$ & $100.0 \pm 6.3$ &
        $100.0 \pm 2.2$ & $100.0 \pm 4.5$ &
        $100.0 \pm 2.1$ & $100.0 \pm 4.4$ 
\\
$[{\rm O\;III}]\;4959$ &
        $87.8 \pm 5.4$ & $87 \pm 12$ &
        $79.7 \pm 4.8$ & $79.0 \pm 9.1$ &
        $69.4 \pm 5.2$ & $68.7 \pm 8.8$ 
\\ 
$[{\rm O\;III}]\;5007$ &
        $260.2 \pm 6.6$ & $258 \pm 28$ &
        $192.9 \pm 5.8$ & $191 \pm 18$ &
        $159.8 \pm 6.1$ & $158 \pm 15$ 
\\
\tableline 
& \multicolumn{2}{c}{Ho II--4} & 
\multicolumn{2}{c}{Ho II--5} &
\multicolumn{2}{c}{Ho II--6} \\
\cline{2-7}
\multicolumn{1}{c}{Identification (\AA)} &
\multicolumn{1}{c}{$F$} & \multicolumn{1}{c}{$I$} &
\multicolumn{1}{c}{$F$} & \multicolumn{1}{c}{$I$} &
\multicolumn{1}{c}{$F$} & \multicolumn{1}{c}{$I$} \\ 
\tableline
$[\rm{O\;II}]\;3727$ &
        $251.1 \pm 8.1$ & $246 \pm 25$ & 
        $191.4 \pm 2.6$ & $188 \pm 15$ &
        $261.3 \pm 3.2$ & $257 \pm 20$ 
\\ 
$[\rm{Ne\;III}]\;3869$ &
        \nodata & \nodata &
        $26.7 \pm 1.2$ & $26.2 \pm 2.6$ &
        $23.2 \pm 1.9$ & $22.8 \pm 3.0$ 
\\
${\rm He\;I} + {\rm H}8\;3889$ &
        \nodata & \nodata &
        $17.2 \pm 1.1$ & $21.1 \pm 2.7$ &
        $21.0 \pm 1.9$ & $24.3 \pm 3.8$
\\ 
$[{\rm Ne\;III}] + {\rm H}\epsilon\;3970$ &
        $18.7 \pm 3.9$ & $21.7 \pm 6.5$ &
        $21.9 \pm 1.6$ & $25.2 \pm 3.3$ &
        $21.5 \pm 2.6$ & $24.6 \pm 4.5$
\\ 
${\rm H}\delta\;4101$ &
        $20.2 \pm 3.8$ & $22.7 \pm 6.2$ &
        $24.3 \pm 1.5$ & $27.4 \pm 3.3$ &
        $29.7 \pm 2.2$ & $32.2 \pm 4.2$
\\ 
${\rm H}\gamma\;4340$ &
        $50.0 \pm 2.7$ & $51.5 \pm 6.3$ &
        $47.72 \pm 0.94$ & $49.8 \pm 4.1$ &
        $43.8 \pm 2.4$ & $45.6 \pm 5.1$
\\ 
$[{\rm O\;III}]\;4363$ &
        $<$ 4.7 (2$\sigma$) & $<$ 3.5 (2$\sigma$) &
        $6.02 \pm 0.75$ & $5.9 \pm 1.0$ &
        $<$ 3.0 (2$\sigma$) & $<$ 2.2 (2$\sigma$) 
\\ 
${\rm He\;I}\;4472$ &
        \nodata & \nodata &
        $2.65 \pm 0.91$ & $2.6 \pm 1.0$ &
        $4.1 \pm 1.2$ & $4.0 \pm 1.3$
\\ 
${\rm He\;II}\;4686$ &
        \nodata & \nodata &
        $2.90 \pm 0.59$ & $2.85 \pm 0.71$ &
        \nodata & \nodata
\\ 
${\rm H}\beta\;4861$ &
        $100.0 \pm 3.2$ & $100.0 \pm 5.2$ &
        $100.0 \pm 1.2$ & $100.0 \pm 4.0$ &
        $100.0 \pm 1.2$ & $100.0 \pm 4.0$
\\ 
$[{\rm O\;III}]\;4959$ &
        $41.9 \pm 3.3$ & $41.1 \pm 5.7$ &
        $96.0 \pm 2.2$ & $94.1 \pm 7.7$ &
        $72.7 \pm 1.8$ & $72.5 \pm 6.0$
\\ 
$[{\rm O\;III}]\;5007$ &
        $139.2 \pm 4.1$ & $137 \pm 14$ &
        $269.5 \pm 2.7$ & $264 \pm 21$ &
        $213.5 \pm 2.2$ & $211 \pm 16$
\\ 
\tableline 
& \multicolumn{2}{c}{Ho II--7} & 
\multicolumn{2}{c}{Ho II--8} &
\multicolumn{2}{c}{Ho II--9} \\
\cline{2-7}
\multicolumn{1}{c}{Identification (\AA)} &
\multicolumn{1}{c}{$F$} & \multicolumn{1}{c}{$I$} &
\multicolumn{1}{c}{$F$} & \multicolumn{1}{c}{$I$} &
\multicolumn{1}{c}{$F$} & \multicolumn{1}{c}{$I$} \\ 
\tableline
$[{\rm O\;II}]\;3727$ &
        $239.2 \pm 2.1$ & $235 \pm 18$ & 
        $221.0 \pm 2.0$ & $217 \pm 17$ &
        $144.8 \pm 2.2$ & $143 \pm 12$ 
\\
$[{\rm Ne\;III}]\;3869$ &
        $23.4 \pm 1.1$ & $23.0 \pm 2.3$ &
        $22.5 \pm 1.2$ & $22.1 \pm 2.3$ &
        $18.9 \pm 1.1$ & $18.6 \pm 2.1$
\\
${\rm He\;I} + {\rm H}8\;3889$ &
        $20.3 \pm 1.1$ & $22.8 \pm 2.5$ &
        $19.5 \pm 1.2$ & $21.4 \pm 2.6$ &
        $15.1 \pm 1.1$ & $16.7 \pm 2.2$ 
\\
$[{\rm Ne III}] + {\rm H}\epsilon\;3970$ &
        $22.2 \pm 1.7$ & $24.6 \pm 3.3$ &
        $22.4 \pm 1.5$ & $24.2 \pm 3.0$ &
        $17.54 \pm 0.70$ & $19.1 \pm 1.9$
\\
${\rm H}\delta\;4101$ &
        $27.3 \pm 1.2$ & $29.5 \pm 3.0$ &
        $25.5 \pm 1.1$ & $27.5 \pm 2.7$ &
        $21.56 \pm 0.73$ & $23.2 \pm 2.2$
\\
${\rm H}\gamma\;4340$ &
        $47.3 \pm 1.4$ & $48.8 \pm 4.3$ &
        $49.5 \pm 1.2$ & $50.8 \pm 4.3$ &
        $42.85 \pm 0.86$ & $43.9 \pm 3.7$
\\
$[{\rm O\;III}]\;4363$ &
        $3.2 \pm 1.1$ & $3.2 \pm 1.2$ &
        $4.30 \pm 0.92$ & $4.2 \pm 1.1$ &
        $5.69 \pm 0.69$ & $5.61 \pm 0.94$
\\
${\rm He\;I}\;4472$ &
        $2.95 \pm 0.90$ & $2.9 \pm 1.0$ &
        $3.43 \pm 0.90$ & $3.4 \pm 1.0$ &
        $3.11 \pm 0.64$ & $3.07 \pm 0.77$
\\
${\rm H}\beta\;4861$ &
        $100.00 \pm 0.91$ & $100.0 \pm 3.9$ &
        $100.0 \pm 1.0$ & $100.0 \pm 3.9$ &
        $100.0 \pm 1.4$ & $100.0 \pm 4.0$
\\
$[{\rm O III}]\;4959$ &
        $73.7 \pm 1.2$ & $72.5 \pm 5.7$ &
        $75.0 \pm 1.3$ & $73.6 \pm 5.9$ &
        $105.3 \pm 3.9$ & $103.9 \pm 9.6$
\\
$[{\rm O III}]\;5007$ &
        $215.1 \pm 1.5$ & $211 \pm 16$ &
        $217.6 \pm 1.5$ & $214 \pm 16$ &
        $300.7 \pm 4.8$ & $297 \pm 24$
\\ 
\tableline
\end{tabular}
\tablecomments{
Wavelengths are listed in \AA.
$F$ is the observed flux ratio with respect to \protect \hbeta.
$I$ is the corrected intensity ratio, corrected only for underlying
Balmer absorption.
The reddening was assumed to be zero.
The errors in the observed line ratios account for the errors in
the fits to the line profiles, the surrounding continua, and
the relative error in the sensitivity function listed in
Table~\protect \ref{table_obslog}.
Flux errors in the \protect \hbeta\ reference line are not included
in the ratio relative to \hbeta.
Errors in the corrected line ratios account for errors in the
specified line and in the \protect \hbeta\ reference line.
}
\end{center}
\end{table}


\begin{table}
\footnotesize 
\begin{center}
\renewcommand{\arraystretch}{1.}
\caption{
Observed and corrected line ratios for IC 10 at OAN.
\vspace*{3mm}
\label{table_ic10_oan}
}
\begin{tabular}{ccccccc}
\tableline \tableline
& \multicolumn{2}{c}{IC 10--1} & 
\multicolumn{2}{c}{IC 10--2} &
\multicolumn{2}{c}{IC 10--3} \\
\cline{2-7}
\multicolumn{1}{c}{Identification (\AA)} &
\multicolumn{1}{c}{$F$} & \multicolumn{1}{c}{$I$} &
\multicolumn{1}{c}{$F$} & \multicolumn{1}{c}{$I$} &
\multicolumn{1}{c}{$F$} & \multicolumn{1}{c}{$I$} \\ 
\tableline
${\rm H}\beta\;4861$ &
        $100.0 \pm 2.2$ & $100.0 \pm 3.8$ &
        $100.0 \pm 1.7$ & $100.0 \pm 3.5$ &
        $100.0 \pm 3.9$ & $100.0 \pm 5.1$
\\
${\rm H}\alpha\;6563$ &
        $628.0 \pm 4.0$ & $286 \pm 20$ &
        $620.0 \pm 3.4$ & $286 \pm 19$ &
        $676.0 \pm 5.3$ & $286 \pm 24$
\\
\multicolumn{1}{c}{$E(B-V)$ (mag)} &
\multicolumn{2}{c}{$+0.788 \pm 0.070$} &
\multicolumn{2}{c}{$+0.773 \pm 0.067$} &
\multicolumn{2}{c}{$+0.862 \pm 0.083$} 
\\
\tableline
& \multicolumn{2}{c}{IC 10--4} & 
\multicolumn{2}{c}{IC 10--5} &
\multicolumn{2}{c}{} \\
\cline{2-5}
\multicolumn{1}{c}{Identification (\AA)} &
\multicolumn{1}{c}{$F$} & \multicolumn{1}{c}{$I$} &
\multicolumn{1}{c}{$F$} & \multicolumn{1}{c}{$I$} &
\multicolumn{1}{c}{} & \multicolumn{1}{c}{} \\ 
\cline{1-5}
${\rm H}\beta\;4861$ &
        $100.0 \pm 4.5$ & $100.0 \pm 5.7$ &
        $100.0 \pm 5.3$ & $100.0 \pm 6.3$ &
        &
\\
${\rm H}\alpha\;6563$ &
        $703.0 \pm 7.7$ & $286 \pm 26$ &
        $961.0 \pm 8.6$ & $286 \pm 27$ &
        &
\\
\multicolumn{1}{c}{$E(B-V)$ (mag)} &
\multicolumn{2}{c}{$+0.897 \pm 0.091$} &
\multicolumn{2}{c}{$+1.215 \pm 0.096$} &
\multicolumn{2}{c}{} 
\\
\cline{1-5}
\end{tabular}
\tablecomments{
Wavelengths are listed in \AA.
$F$ is the observed flux ratio with respect to \protect \hbeta.
$I$ is the intensity ratio corrected for underlying Balmer absorption
and for reddening.
}
\end{center}
\end{table}


\begin{table}
\footnotesize
\begin{center}
\renewcommand{\arraystretch}{1.}
\caption{
Observed and corrected line ratios for IC 10 at Steward.
\vspace*{3mm}
\label{table_ic10_data}
}
\begin{tabular}{ccccccc}
\tableline \tableline
& \multicolumn{2}{c}{IC 10--1} & 
\multicolumn{2}{c}{IC 10--2} &
\multicolumn{2}{c}{IC 10--3} \\
\cline{2-7}
\multicolumn{1}{c}{Identification (\AA)} &
\multicolumn{1}{c}{$F$} & \multicolumn{1}{c}{$I$} &
\multicolumn{1}{c}{$F$} & \multicolumn{1}{c}{$I$} &
\multicolumn{1}{c}{$F$} & \multicolumn{1}{c}{$I$} \\ 
\tableline
$[\rm{O\;II}]\;3727$ &
        $68.5 \pm 1.1$ & $198 \pm 30$ &
        $61.9 \pm 1.3$ & $175 \pm 26$ &
        $98.7 \pm 2.6$ & $312 \pm 56$ 
\\
$[\rm{Ne\;III}]\;3869$ &
        $12.28 \pm 0.70$ & $29.6 \pm 5.0$ &
        $12.64 \pm 0.69$ & $30.0 \pm 4.9$ &
        $12.8 \pm 1.9$ & $33.2 \pm 9.0$
\\ 
${\rm He\;I} + {\rm H8}\;3889$ &
        $8.64 \pm 0.67$ & $25.8 \pm 5.2$ &
        $9.36 \pm 0.66$ & $26.5 \pm 5.0$ &
        $7.5 \pm 1.5$ & $27 \pm 11$
\\ 
$[{\rm Ne\;III}] + {\rm H}\epsilon\;3970$ &
        $10.99 \pm 0.63$ & $28.4 \pm 4.8$ &
        $11.21 \pm 0.67$ & $28.1 \pm 4.7$ &
        $10.6 \pm 1.2$ & $31.1 \pm 8.0$
\\ 
${\rm H}\delta\;4101$ &
        $15.32 \pm 0.49$ & $32.6 \pm 4.2$ &
        $15.49 \pm 0.49$ & $32.3 \pm 4.0$ &
        $15.5 \pm 1.2$ & $35.8 \pm 6.7$
\\
${\rm H}\gamma\;4340$ &
        $34.18 \pm 0.40$ & $53.5 \pm 4.8$ &
        $35.28 \pm 0.41$ & $54.5 \pm 4.7$ &
        $29.36 \pm 0.83$ & $48.9 \pm 5.4$
\\
$[{\rm O\;III}]\;4363$ &
        $1.65 \pm 0.32$ & $2.37 \pm 0.60$ &
        $1.53 \pm 0.33$ & $2.20 \pm 0.60$ &
        $2.77 \pm 0.73$ & $4.1 \pm 1.4$
\\ 
${\rm He\;I}\;4472$ &
        $3.48 \pm 0.31$ & $4.55 \pm 0.66$ &
        $3.46 \pm 0.32$ & $4.52 \pm 0.66$ &
        $3.71 \pm 0.71$ & $4.9 \pm 1.3$
\\
${\rm He\;II}\;4686$ &
        \nodata & \nodata & \nodata & \nodata &
        $1.44 \pm 0.53$ & $1.60 \pm 0.65$
\\ 
${\rm H}\beta\;4861$ &
        $100.0 \pm 1.3$ & $100.0 \pm 2.7$ &
        $100.0 \pm 1.3$ & $100.0 \pm 2.6$ &
        $100.0 \pm 1.7$ & $100.0 \pm 2.9$
\\ 
$[{\rm O\;III}]\;4959$ &
        $134.5 \pm 5.9$ & $123.4 \pm 8.6$ &
        $136.7 \pm 5.7$ & $126.9 \pm 8.6$ &
        $127.7 \pm 6.3$ & $115.3 \pm 8.8$
\\ 
$[{\rm O\;III}]\;5007$ &
        $404.8 \pm 7.4$ & $360 \pm 17$ &
        $409.1 \pm 7.2$ & $366 \pm 17$ &
        $391.6 \pm 7.9$ & $342 \pm 17$
\\ 
\tableline
& \multicolumn{2}{c}{IC 10--4} & 
\multicolumn{2}{c}{IC 10--5} &
\multicolumn{2}{c}{} \\
\cline{2-5}
\multicolumn{1}{c}{Identification (\AA)} &
\multicolumn{1}{c}{$F$} & \multicolumn{1}{c}{$I$} &
\multicolumn{1}{c}{$F$} & \multicolumn{1}{c}{$I$} &
\multicolumn{1}{c}{} & \multicolumn{1}{c}{} \\ 
\cline{1-5}
$[\rm{O\;II}]\;3727$ &
        $154.2 \pm 2.9$ & $517 \pm 98$ &
        $105.0 \pm 6.2$ & $550 \pm 130$ & &
\\ 
$[\rm{Ne\;III}]\;3869$ &
	\nodata & \nodata & \nodata & \nodata & & \\ 
${\rm He\;I} + {\rm H8}\;3889$ &
	$9.2 \pm 1.5$ & $29.4 \pm 9.8$ & \nodata & \nodata & &
\\
$[{\rm Ne\;III}] + {\rm H}\epsilon\;3970$ &
        $12.6 \pm 2.1$ & $34 \pm 11$ &
        $12.3 \pm 2.8$ & $44 \pm 17$ & &
\\ 
${\rm H}\delta\;4101$ &
        $10.9 \pm 1.3$ & $26.5 \pm 6.8$ &
        $20.4 \pm 2.6$ & $56 \pm 15$ & &
\\ 
${\rm H}\gamma\;4340$ &
        $32.0 \pm 1.3$ & $52.9 \pm 6.9$ &
        $29.0 \pm 2.6$ & $55 \pm 11$ & &
\\
$[{\rm O\;III}]\;4363$ &
        $<$ 1.6 (2$\sigma$) & $<$ 2.5 (2$\sigma$) &
        $<$ 4.8 (2$\sigma$) & $<$ 8.7 (2$\sigma$) & &
\\
${\rm He\;I}\;4472$ &
	$2.04 \pm 0.60$ & $2.8 \pm 1.0$ & \nodata & \nodata & & 
\\
${\rm He\;II}\;4686$ &
	\nodata & \nodata & \nodata & \nodata & & \\
${\rm H}\beta\;4861$ &
        $100.0 \pm 2.5$ & $100.0 \pm 3.5$ &
        $100.0 \pm 3.6$ & $100.0 \pm 4.5$ & &
\\
$[{\rm O\;III}]\;4959$ &
        $63.3 \pm 4.3$ & $57.6 \pm 5.7$ &
        $102.9 \pm 9.2$ & $92 \pm 12$ & & 
\\
$[{\rm O\;III}]\;5007$ &
        $194.8 \pm 5.4$ & $171 \pm 10$ &
        $326.1 \pm 11.8$ & $278 \pm 21$ & &
\\
\tableline
\end{tabular}
\tablecomments{
Wavelengths are listed in \AA.
$F$ is the observed flux ratio with respect to \hbeta.
$I$ is the corrected intensity ratio, corrected for underlying
Balmer absorption and for the reddening value of each \hii\ region
derived from OAN data in the previous table.
The reddening was assumed to be zero.
The errors in the observed line ratios account for the errors in
the fits to the line profiles, the surrounding continua, and
the relative error in the sensitivity function listed in
Table~\ref{table_obslog}.
Flux errors in the \hbeta\ reference line are not included 
in the ratio relative to \hbeta.
Errors in the corrected line ratios account for errors in the
specified line and in the \hbeta\ reference line.
}
\end{center}
\end{table}


\begin{table}
\footnotesize 
\begin{center}
\renewcommand{\arraystretch}{1.}
\caption{
Observed and corrected line ratios for NGC 1560 (red spectra).
\vspace*{3mm}
\label{table_n1560_reddata}
}
\begin{tabular}{ccccccc}
\tableline \tableline
& \multicolumn{2}{c}{NGC 1560--1 NE} & 
\multicolumn{2}{c}{NGC 1560--2 NE} &
\multicolumn{2}{c}{} \\
\cline{2-5}
\multicolumn{1}{c}{Identification (\AA)} &
\multicolumn{1}{c}{$F$} & \multicolumn{1}{c}{$I$} &
\multicolumn{1}{c}{$F$} & \multicolumn{1}{c}{$I$} &
\multicolumn{1}{c}{} & \multicolumn{1}{c}{} \\ 
\cline{1-5}
${\rm H}\beta\;4861$ &
        $100.0 \pm 6.3$ & $100.0 \pm 7.0$ &
        $100 \pm 21$ & $100 \pm 24$ &
        & 
\\
${\rm H}\alpha\;6563$ &
        $396 \pm 14$ & $286 \pm 44$ &
        $183 \pm 10$ & $286 \pm 86$ &
        &
\\
\multicolumn{1}{c}{$E(B-V)$ (mag)} &
\multicolumn{2}{c}{$+0.33 \pm 0.15$} &
\multicolumn{2}{c}{$-0.47 \pm 0.29$} &
\multicolumn{2}{c}{} 
\\
\tableline
& \multicolumn{2}{c}{NGC 1560--1 SW} & 
\multicolumn{2}{c}{NGC 1560--2 SW} &
\multicolumn{2}{c}{NGC 1560--3 SW} \\
\cline{2-7}
\multicolumn{1}{c}{Identification (\AA)} &
\multicolumn{1}{c}{$F$} & \multicolumn{1}{c}{$I$} &
\multicolumn{1}{c}{$F$} & \multicolumn{1}{c}{$I$} &
\multicolumn{1}{c}{$F$} & \multicolumn{1}{c}{$I$} \\ 
\cline{1-7}
${\rm H}\beta\;4861$ &
        $100.0 \pm 8.3$ & $100 \pm 10$ &
        $100.0 \pm 8.2$ & $100 \pm 10$ &
        $100.0 \pm 8.5$ & $100 \pm 10$ 
\\
${\rm H}\alpha\;6563$ &
        $523 \pm 15$ & $286 \pm 39$ &
        $465 \pm 15$ & $286 \pm 40$ &
        $397 \pm 17$ & $286 \pm 43$ 
\\ 
\multicolumn{1}{c}{$E(B-V)$ (mag)} &
\multicolumn{2}{c}{$+0.48 \pm 0.14$} &
\multicolumn{2}{c}{$+0.38 \pm 0.14$} &
\multicolumn{2}{c}{$+0.24 \pm 0.15$} 
\\
\tableline
& \multicolumn{2}{c}{NGC 1560--6 SW} & 
\multicolumn{2}{c}{NGC 1560--7 SW} &
\multicolumn{2}{c}{} \\
\cline{2-5}
\multicolumn{1}{c}{Identification (\AA)} &
\multicolumn{1}{c}{$F$} & \multicolumn{1}{c}{$I$} &
\multicolumn{1}{c}{$F$} & \multicolumn{1}{c}{$I$} &
\multicolumn{1}{c}{} & \multicolumn{1}{c}{} \\ 
\cline{1-5}
${\rm H}\beta\;4861$ &
        $100.0 \pm 7.9$ & $100.0 \pm 8.9$ & 
        $100 \pm 34$ & $100 \pm 46$ & 
        & 
\\ 
${\rm H}\alpha\;6563$ &
        $445 \pm 13$ & $286 \pm 36$ & 
        $556 \pm 15$ & $286 \pm 155$ & 
        & 
\\ 
\multicolumn{1}{c}{$E(B-V)$ (mag)} &
\multicolumn{2}{c}{$+0.38 \pm 0.13$} &
\multicolumn{2}{c}{$+0.56 \pm 0.51$} &
\multicolumn{2}{c}{} 
\\
\cline{1-5}
\end{tabular}
\tablecomments{
Wavelengths are listed in \AA.
$F$ is the observed flux ratio with respect to \protect \hbeta.
$I$ is the intensity ratio corrected for underlying Balmer absorption
and for reddening.
The reddening adopted for NGC~1560 is the average of the reddening
values for \hii\ regions 1~NE, 1~SW, 2~SW, 3~SW, and 6~SW.
}
\end{center}
\end{table}

\clearpage 

\begin{table}
\scriptsize
\begin{center}
\renewcommand{\arraystretch}{1.}
\caption{
Observed and corrected line ratios for NGC 1560 (blue spectra).
\vspace*{3mm}
\label{table_n1560_data}
}
\begin{tabular}{ccccccc}
\tableline \tableline
& \multicolumn{2}{c}{NGC 1560--1} & 
\multicolumn{2}{c}{NGC 1560--2} &
\multicolumn{2}{c}{NGC 1560--3} \\
\cline{2-7}
\multicolumn{1}{c}{Identification (\AA)} &
\multicolumn{1}{c}{$F$} & \multicolumn{1}{c}{$I$} &
\multicolumn{1}{c}{$F$} & \multicolumn{1}{c}{$I$} &
\multicolumn{1}{c}{$F$} & \multicolumn{1}{c}{$I$} \\ 
\tableline
$[\rm{O\;II}]\;3727$ &
        $154.3 \pm 6.7$ & $254 \pm 22$ &
        $224.7 \pm 4.8$ & $359 \pm 23$ &
        $209\pm 20$ & $338 \pm 56$
\\
$[\rm{Ne\;III}]\;3869$ &
        $8.9 \pm 3.8$ & $13.5 \pm 6.2$ &
        $13.0 \pm 2.1$ & $19.2 \pm 3.8$ &
        $31.8 \pm 7.2$ & $48 \pm 14$ 
\\ 
${\rm He\;I} + {\rm H8}\;3889$ &
        \nodata & \nodata &
        $19.2 \pm 2.4$ & $31.6 \pm 5.6$ &
        \nodata & \nodata
\\ 
$[{\rm Ne\;III}] + {\rm H}\epsilon\;3970$ &
        $12.2 \pm 2.7$ & $17.0 \pm 5.2$ &
        $10.7 \pm 1.8$ & $18.8 \pm 4.5$ &
        \nodata & \nodata
\\
${\rm H}\delta\;4101$ &
        $31.4 \pm 4.9$ & $40.9 \pm 8.4$ &
        $21.0 \pm 1.7$ & $30.8 \pm 4.0$ &
        $26.2 \pm 5.6$ & $38 \pm 12$
\\ 
${\rm H}\gamma\;4340$ &
        $40.7 \pm 2.1$ & $48.5 \pm 4.6$ &
        $43.9 \pm 1.5$ & $54.4 \pm 4.3$ &
        $50.0 \pm 5.0$ & $61 \pm 11$
\\ 
$[{\rm O\;III}]\;4363$ &
        $<$ 2.8 (2$\sigma$) & $<$ 3.4 (2$\sigma$) &
        $<$ 2.8 (2$\sigma$) & $<$ 3.3 (2$\sigma$) &
        $<$ 11.5 (2$\sigma$) & $<$ 13.6 (2$\sigma$) 
\\
${\rm He\;I}\;4472$ &
        \nodata & \nodata &
        $4.02 \pm 0.75$ & $4.5 \pm 1.0$ &
        \nodata & \nodata
\\ 
${\rm H}\beta\;4861$ &
        $100.0 \pm 2.6$ & $100.0 \pm 3.6$ &
        $100.0 \pm 2.4$ & $100.0 \pm 3.4$ &
        $100.0 \pm 5.6$ & $100.0 \pm 7.0$
\\ 
$[{\rm O\;III}]\;4959$ &
        $124.6 \pm 4.7$ & $122.3 \pm 9.8$ &
        $81.8 \pm 5.2$ & $78.1 \pm 8.0$ &
        $150 \pm 11$ & $145 \pm 21$
\\ 
$[{\rm O\;III}]\;5007$ &
        $324.8 \pm 5.8$ & $315 \pm 20$ &
        $238.1 \pm 6.6$ & $224 \pm 16$ &
        $426 \pm 14$ & $407 \pm 44$
\\ 
\tableline
& \multicolumn{2}{c}{NGC 1560--4} & 
\multicolumn{2}{c}{NGC 1560--5} &
\multicolumn{2}{c}{NGC 1560--6} \\
\cline{2-7}
\multicolumn{1}{c}{Identification (\AA)} &
\multicolumn{1}{c}{$F$} & \multicolumn{1}{c}{$I$} &
\multicolumn{1}{c}{$F$} & \multicolumn{1}{c}{$I$} &
\multicolumn{1}{c}{$F$} & \multicolumn{1}{c}{$I$} \\ 
\cline{1-7}
$[\rm{O\;II}]\;3727$ &
        $220 \pm 18$ & $363 \pm 49$ &
        $254 \pm 18$ & $423 \pm 56$ &
        $295 \pm 37$ & $385 \pm 91$ 
\\ 
$[\rm{Ne\;III}]\;3869$ &
        $19.8 \pm 6.9$ & $30 \pm 12$ & 
        \nodata & \nodata &
        \nodata & \nodata
\\
${\rm He\;I} + {\rm H8}\;3889$ &
        \nodata & \nodata & 
        $38.7 \pm 5.9$ & $58 \pm 13$ &
        \nodata & \nodata
\\ 
$[{\rm Ne\;III}] + {\rm H}\epsilon\;3970$ &
        $23.2 \pm 5.3$ & $33.9 \pm 9.0$ &
        $27.8 \pm 6.7$ & $39 \pm 13$ &
        \nodata & \nodata
\\
${\rm H}\delta\;4101$ &
        $34.8 \pm 6.1$ & $46 \pm 11$ &
        $29.3 \pm 5.7$ & $38 \pm 10$ &
        \nodata & \nodata
\\ 
${\rm H}\gamma\;4340$ &
        $52.1 \pm 3.9$ & $61.3 \pm 8.4$ &
        $44.5 \pm 4.1$ & $51.6 \pm 8.4$ &
        \nodata & \nodata
\\ 
$[{\rm O\;III}]\;4363$ &
        $<$ 6.7 (2$\sigma$) & $<$ 8.1 (2$\sigma$) &
        $<$ 4.4 (2$\sigma$) & $<$ 5.3 (2$\sigma$) &
        \nodata & \nodata
\\ 
${\rm H}\beta\;4861$ &
        $100.0 \pm 4.5$ & $100.0 \pm 5.3$ &
        $100.0 \pm 5.2$ & $100.0 \pm 6.0$ &
        $100.0 \pm 8.5$ & $100 \pm 11$ 
\\
$[{\rm O\;III}]\;4959$ &
        $110.5 \pm 8.2$ & $109 \pm 14$ &
        $58.5 \pm 4.9$ & $58.3 \pm 8.5$ &
        $40.3 \pm 6.8$ & $31.4 \pm 8.7$ 
\\
$[{\rm O\;III}]\;5007$ &
        $318 \pm 11$ & $310 \pm 29$ &
        $175.8 \pm 6.2$ & $173 \pm 18$ &
	$130.1 \pm 9.4$ & $100 \pm 18$ 
\\
\tableline
& \multicolumn{2}{c}{NGC 1560--7} & 
\multicolumn{2}{c}{} &
\multicolumn{2}{c}{} \\
\cline{2-3}
\multicolumn{1}{c}{Identification (\AA)} &
\multicolumn{1}{c}{$F$} & \multicolumn{1}{c}{$I$} &
\multicolumn{1}{c}{} & \multicolumn{1}{c}{} &
\multicolumn{1}{c}{} & \multicolumn{1}{c}{} \\ 
\cline{1-3}
$[\rm{O\;II}]\;3727$ &
	$182.1 \pm 67.2$ & $240 \pm 140$ & & & & 
\\
${\rm H}\beta\;4861$ &
	$100.0 \pm 16.2$ & $100 \pm 20$ & & & & 
\\
$[{\rm O\;III}]\;4959$ &
	$37.1 \pm 17.7$ & $29 \pm 20$ & & & &
\\
$[{\rm O\;III}]\;5007$ &
	$214.02 \pm 23.2$ & $163 \pm 52$ & & & & 
\\
\tableline
\end{tabular}
\tablecomments{
Wavelengths are listed in \AA.
$F$ is the observed flux ratio with respect to \hbeta.
$I$ is the corrected intensity ratio, corrected for underlying
Balmer absorption and for the average reddening of 
$E(B-V) = +0.36$ mag derived from OAN data.
The errors in the observed line ratios account for the errors in
the fits to the line profiles, the surrounding continua, and
the relative error in the sensitivity function listed in
Table~\ref{table_obslog}.
Flux errors in the \protect \hbeta\ reference line are not included
in the ratio relative to \hbeta.
Errors in the corrected line ratios account for errors in the
specified line and in the \hbeta\ reference line.
}
\end{center}
\end{table}


\begin{table}
\scriptsize
\begin{center}
\renewcommand{\arraystretch}{1.}
\caption{
Observed and corrected line ratios for NGC~3109.
\vspace*{3mm}
\label{table_n3109_data}
}
\begin{tabular}{ccccc}
\tableline \tableline
& \multicolumn{2}{c}{NGC 3109--1} & 
\multicolumn{2}{c}{NGC 3109--2} \\
\cline{2-5}
\multicolumn{1}{c}{Identification (\AA)} &
\multicolumn{1}{c}{$F$} & \multicolumn{1}{c}{$I$} &
\multicolumn{1}{c}{$F$} & \multicolumn{1}{c}{$I$} \\ 
\tableline
$[\rm{O\;II}]\;3727$ &
        $418 \pm 77$ & $420 \pm 200$ &
        $385 \pm 11$ & $380 \pm 44$ \\ 
$[\rm{Ne\;III}]\;3869$ &
	\nodata & \nodata & \nodata & \nodata \\ 
${\rm He\;I} + {\rm H8}\;3889$ &
	\nodata & \nodata & \nodata & \nodata \\ 
$[{\rm Ne\;III}] + {\rm H}\epsilon\;3970$ &
	\nodata & \nodata & \nodata & \nodata \\ 
${\rm H}\delta\;4101$ &
	\nodata & \nodata & \nodata & \nodata \\
${\rm H}\gamma\;4340$ &
        \nodata & \nodata & $42.3 \pm 5.6$ & $45 \pm 12$ \\
$[{\rm O\;III}]\;4363$ &
	\nodata & \nodata & \nodata & \nodata \\ 
${\rm H}\beta\;4861$ &
        $100 \pm 26$ & $100 \pm 28$ &
        $100.0 \pm 5.2$ & $100 \pm 6.8$ \\
$[{\rm O\;III}]\;4959$ &
        $41 \pm 23$ & $41 \pm 36$ & \nodata & \nodata \\
$[{\rm O\;III}]\;5007$ &
        $202 \pm 30$ & $204 \pm 90$ &
        $24.2 \pm 4.6$ & $23.9 \pm 6.2$ \\
\tableline
& \multicolumn{2}{c}{NGC 3109--3} & 
\multicolumn{2}{c}{NGC 3109--4} \\
\cline{2-5}
\multicolumn{1}{c}{Identification (\AA)} &
\multicolumn{1}{c}{$F$} & \multicolumn{1}{c}{$I$} &
\multicolumn{1}{c}{$F$} & \multicolumn{1}{c}{$I$} \\ 
\tableline
$[\rm{O\;II}]\;3727$ &
        $227.5 \pm 6.2$ & $220 \pm 20$ &
        $304.8 \pm 7.4$ & $283 \pm 26$ \\ 
$[\rm{Ne\;III}]\;3869$ &
        $10.9 \pm 2.1$ & $10.5 \pm 2.5$ & \nodata & \nodata \\
${\rm He\;I} + {\rm H8}\;3889$ &
        $16.5 \pm 2.8$ & $21.1 \pm 5.6$ & \nodata & \nodata \\ 
$[{\rm Ne\;III}] + {\rm H}\epsilon\;3970$ &
        $16.7 \pm 3.2$ & $20.9 \pm 6.1$ & \nodata & \nodata \\
${\rm H}\delta\;4101$ &
        $20.9 \pm 3.6$ & $25.0 \pm 6.4$ &
        $17.3 \pm 2.7$ & $27.1 \pm 7.6$ \\
${\rm H}\gamma\;4340$ &
        $45.3 \pm 2.4$ & $47.7 \pm 5.5$ &
        $41.1 \pm 3.2$ & $47.4 \pm 7.1$ \\
$[{\rm O\;III}]\;4363$ &
        $5.7 \pm 2.0$ & $5.5 \pm 2.2$ &
        $<$ 6.5 (2$\sigma$) & $<$ 6.0 (2$\sigma$) \\
${\rm H}\beta\;4861$ &
        $100.0 \pm 2.2$ & $100.0 \pm 4.4$ &
        $100.0 \pm 2.5$ & $100.0 \pm 4.7$ \\
$[{\rm O\;III}]\;4959$ &
        $105.7 \pm 3.3$ & $102.3 \pm 9.5$ &
        $100.6 \pm 3.4$ & $93.3 \pm 9.1$ \\
$[{\rm O\;III}]\;5007$ &
        $313.4 \pm 4.1$ & $303 \pm 25$ &
        $280.6 \pm 4.1$ & $260 \pm 23$ \\ 
\tableline
\end{tabular}
\tablecomments{
Wavelengths are listed in \AA.
$F$ is the observed flux ratio with respect to \hbeta\ $\equiv 100$
and $I$ is the intensity ratio corrected only for underlying Balmer
absorption.
The reddening was assumed to be zero.
The errors in the observed line ratios account only for the errors in
the fits to the line profiles and surrounding continua, and
the relative error in the sensitivity function
(Table~\ref{table_obslog}).
Flux errors in the \hbeta\ reference line are not included
in the ratio relative to \hbeta.
However, errors in the corrected line ratios account for flux
errors in both the specified line and in the reference line.   
}
\end{center}
\end{table}


\begin{table}
\footnotesize
\begin{center}
\renewcommand{\arraystretch}{1.}
\caption{
Observed and corrected line ratios for UGC~6456.
\vspace*{3mm}
\label{table_u6456_data}
}
\begin{tabular}{ccccc}
\tableline \tableline
& \multicolumn{2}{c}{UGC 6456--1} & 
\multicolumn{2}{c}{UGC 6456--2} \\
\cline{2-5}
\multicolumn{1}{c}{Identification (\AA)} &
\multicolumn{1}{c}{$F$} & \multicolumn{1}{c}{$I$} &
\multicolumn{1}{c}{$F$} & \multicolumn{1}{c}{$I$} \\ 
\tableline
$[\rm{O\;II}]\;3727$ &
        $136.8 \pm 6.1$ & $133 \pm 16$ &
        $116.2 \pm 1.9$ & $114.6 \pm 9.2$ \\ 
$[\rm{Ne\;III}]\;3869$ &
        $12.5 \pm 2.5$ & $12.1 \pm 3.2$ &
        $25.7 \pm 1.2$ & $25.3 \pm 2.5$ \\ 
${\rm He\;I} + {\rm H8}\;3889$ &
        $12.9 \pm 2.5$ & $18.0 \pm 5.9$ &
        $16.4 \pm 1.1$ & $18.5 \pm 2.3$ \\
$[{\rm Ne\;III}] + {\rm H}\epsilon\;3970$ &
        \nodata & \nodata & $19.9 \pm 1.0$ & $22.0 \pm 2.4$ \\
${\rm H}\delta\;4101$ &
        $21.7 \pm 3.0$ & $25.8 \pm 6.0$ &
        $26.0 \pm 1.1$ & $27.9 \pm 2.9$ \\
${\rm H}\gamma\;4340$ &
        $46.2 \pm 2.7$ & $48.5 \pm 6.6$ &
        $50.2 \pm 1.4$ & $51.3 \pm 4.6$ \\
$[{\rm O\;III}]\;4363$ &
        $8.6 \pm 2.2$ & $8.3 \pm 2.7$ & 
        $8.3 \pm 1.1$ & $8.2 \pm 1.5$ \\
${\rm He\;I}\;4472$ &
        \nodata & \nodata & $4.41 \pm 0.89$ & $4.4 \pm 1.1$ \\
${\rm He\;II}\;4686$ &
        \nodata & \nodata & $1.54 \pm 0.58$ & $1.52 \pm 0.61$ \\ 
${\rm H}\beta\;4861$ &
        $100.0 \pm 4.3$ & $100.0 \pm 6.0$ &
        $100.0 \pm 1.2$ & $100.0 \pm 4.0$ \\
$[{\rm O\;III}]\;4959$ &
        $87.9 \pm 3.3$ & $85.3 \pm 9.7$ &
        $112.0 \pm 3.5$ & $110.5 \pm 9.8$ \\
$[{\rm O\;III}]\;5007$ &
        $237.3 \pm 4.1$ & $231 \pm 23$ &
        $308.5 \pm 4.4$ & $304 \pm 24$ \\
\tableline
\end{tabular}
\tablecomments{
Wavelengths are listed in \AA.
$F$ is the observed flux ratio with respect to \hbeta\ $\equiv 100$
and $I$ is the intensity ratio corrected only for underlying Balmer
absorption.
The reddening was assumed to be zero.
The errors in the observed line ratios account only for the errors in
the fits to the line profiles and surrounding continua, and
the relative error in the sensitivity function
(Table~\ref{table_obslog}).
Flux errors in the \hbeta\ reference line are not included
in the ratio relative to \hbeta.
However, errors in the corrected line ratios account for flux
errors in both the specified line and in the reference line.
}
\end{center}
\end{table}


\begin{table}
\footnotesize
\begin{center}
\renewcommand{\arraystretch}{1.}
\caption{
Observed and corrected line ratios for DDO~187.
\vspace*{3mm}
\label{table_ddo187_data}
}
\begin{tabular}{ccccc}
\tableline \tableline
& \multicolumn{2}{c}{DDO 187--1} & 
\multicolumn{2}{c}{DDO 187--2} \\
\cline{2-5}
\multicolumn{1}{c}{Identification (\AA)} &
\multicolumn{1}{c}{$F$} & \multicolumn{1}{c}{$I$} &
\multicolumn{1}{c}{$F$} & \multicolumn{1}{c}{$I$} \\ 
\tableline
$[\rm{O\;II}]\;3727$ &
        $180 \pm 8.4$ & $175.2 \pm 8.4$ &
        $256 \pm 15$ & $237 \pm 15$ \\ 
$[\rm{Ne\;III}]\;3869$ &
        $27.5 \pm 1.4$ & $26.8 \pm 1.4$ & 
        \nodata & \nodata \\ 
${\rm H}\gamma\;4340$ &
        $44.3 \pm 1.8$ & $44.3 \pm 1.8$ & 
        $35.3 \pm 2.7$ & $35.3 \pm 2.7$ \\ 
$[{\rm O\;III}]\;4363$ &
        $2.7 \pm 0.6$ & $2.6 \pm 0.6$ & 
        \nodata & \nodata \\ 
${\rm H}\beta\;4861$ &
        $100.0 \pm 3.7$ & $100.0 \pm 3.7$ & 
        $100.0 \pm 4.8$ & $100.0 \pm 4.8$ \\ 
$[{\rm O\;III}]\;4959$ &
        $64.5 \pm 2.4$ & $62.8 \pm 2.4$ & 
        $16.6 \pm 2.3$ & $15.4 \pm 2.3$ \\ 
$[{\rm O\;III}]\;5007$ &
        $187.8 \pm 6.8$ & $182.8 \pm 6.8$& 
        $47 \pm 3$ & $43 \pm 3$ \\ 
${\rm H}\alpha\;6563$ &
        $274 \pm 13$ & $274 \pm 13$ & 
        $270 \pm 16$ & $270 \pm 16$ \\ 
$[{\rm N\;II}]\;6583$ &
        $4.3 \pm 0.4$ & $4.2 \pm 0.4$ & 
        $5.7 \pm 0.7$ & $5.3 \pm 0.7$ \\ 
${\rm He\;I}\;6678$ &
        $1.9 \pm 0.3$ & $1.9 \pm 0.3$ &
        \nodata & \nodata \\ 
$[{\rm S\;II}]\;6716$ &
        $78.0 \pm 0.5$ & $75.9 \pm 0.5$ &
        $15.2 \pm 1.1$ & $14.1 \pm 1.1$ \\ 
$[{\rm S\;II}]\;6731$ &
        $63.0 \pm 0.4$ & $61.3 \pm 0.4$ &
        $11.6 \pm 0.9$ & $10.7 \pm 0.9$ \\ 
${\rm He\;I}\;7065$ &
        $1.6 \pm 0.3$ & $1.6 \pm 0.3$ &
        \nodata & \nodata \\ 
$[{\rm Ar\;III}]\;7136$ &
        $3.2 \pm 0.3$ & $3.1 \pm 0.3$ &
        \nodata & \nodata \\ 
\tableline
\end{tabular}
\tablecomments{
Observed and re--analyzed line ratios for DDO 187. 
Wavelengths are listed in \AA.
$F$ is the observed flux ratio with respect to \hbeta\ = 100
\citep{vanzee97}
and $I$ is the intensity ratio corrected for underlying Balmer
absorption with an equivalent width of 2~\AA.
The reddening was assumed to be zero.
}
\end{center}
\end{table}


\begin{deluxetable}{ccccccc}
\renewcommand{\arraystretch}{1.}
\tablecolumns{7}
\tabletypesize{\scriptsize} 
\tablewidth{0pt}
\tablecaption{
Derived properties for \hii\ regions in Holmberg~II, IC~10, 
NGC~1560, NGC~3109, UGC~6456, and DDO~187. 
\label{table_derivedprops}}
\tablehead{
\colhead{} & \colhead{} & \colhead{Derived} & \colhead{Adopted} &
\colhead{} & \colhead{} & \colhead{} \\
\colhead{H II Region} & \colhead{$I($H$\beta)$} &
\colhead{$E(B-V)$} & \colhead{$E(B-V)$} & \colhead{$W_e$(\hbeta)} &
\colhead{$T_e$(O$^{+2}$)} & \colhead{12$+$log(O/H)} \\
& \colhead{(ergs s$^{-1}$ cm$^{-2}$)} & \colhead{(mag)} &
\colhead{(mag)} & \colhead{(\AA)} & \colhead{(K)} & \colhead{(dex)} \\
\colhead{(1)} & \colhead{(2)} & \colhead{(3)} &
\colhead{(4)} & \colhead{(5)} & \colhead{(6)} & \colhead{(7)}
}
\startdata
Ho II--1 & $(1.97 \pm 0.12) \times 10^{-15}$ & 
        $-0.59 \pm 0.29$ & 0 & $209 \pm 79$ & 
	$<$ 17400 & $>$ 7.64 \\
Ho II--2 & $(6.85 \pm 0.31) \times 10^{-15}$ & 
        $-0.46 \pm 0.20$ & 0 & $218 \pm 42$ & 
	$<$ 23500 & $>$ 7.17 \\
Ho II--3 & $(4.47 \pm 0.20) \times 10^{-15}$ & 
        $-0.42 \pm 0.20$ & 0 & $201 \pm 35$ & 
	$<$ 29000 & $>$ 6.89 \\
Ho II--4 & $(3.24 \pm 0.17) \times 10^{-15}$ & 
        $-0.18 \pm 0.24$ & 0 & $104 \pm 14$ & 
	$<$ 17300 & $>$ 7.37 \\
Ho II--5 & $(2.144 \pm 0.085) \times 10^{-14}$ & 
        $-0.12 \pm 0.16$ & 0 & $101.9 \pm 5.0$ & 
	$16000 \pm 1300$ & $7.56 \pm 0.13$ \\
Ho II--6 & $(9.46 \pm 0.37) \times 10^{-15}$ & 
        $+0.05 \pm 0.22$ & 0 & $113.8 \pm 5.8$ &
	$<$ 11700 & $>$ 7.97 \\
Ho II--7 & $(2.099 \pm 0.081) \times 10^{-14}$ & 
        $-0.08 \pm 0.17$ & 0 & $113.5 \pm 4.6$ &
	$13400 \pm 2100$ & $7.76 \pm 0.25$ \\
Ho II--8 & $(1.147 \pm 0.045) \times 10^{-14}$ & 
        $-0.16 \pm 0.17$ & 0 & $112.3 \pm 5.0$ &
	$15100 \pm 1800$ & $7.60 \pm 0.18$ \\
Ho II--9 & $(3.24 \pm 0.13) \times 10^{-14}$ & 
        $+0.12 \pm 0.16$ & 0 & $155 \pm 12$ &
	$14800 \pm 1100$ & $7.65 \pm 0.13$ \\
IC 10--1 & $(1.58 \pm 0.41) \times 10^{-11}$ & 
	$+0.788 \pm 0.070\,$\tablenotemark{a} & $+$0.788 & 
	$88.5 \pm 3.9$ & $10100 \pm 560$ & $8.29 \pm 0.13$ \\
IC 10--2 & $(1.18 \pm 0.29) \times 10^{-11}$ & 
	$+0.773 \pm 0.067\,$\tablenotemark{a} & $+$0.773 &
	$101.0 \pm 4.7$ & $9800 \pm 580$ & 
        $8.32 \pm 0.14\,$\tablenotemark{b} \\
IC 10--3 & $(3.6 \pm 1.1) \times 10^{-12}$ & 
	$+0.862 \pm 0.083\,$\tablenotemark{a} & $+$0.862 &
	$60.5 \pm 2.4$ & $12300 \pm 1100$ & $8.05 \pm 0.18$ \\
IC 10--4 & $(6.3 \pm 2.1) \times 10^{-12}$ & 
	$+0.897 \pm 0.091\,$\tablenotemark{a} & $+$0.897 &
	$92.6 \pm 7.6$ & $<$ 13200 & $>$ 7.96 \\
IC 10--5 & $(1.09 \pm 0.40) \times 10^{-11}$ & 
	$+1.215 \pm 0.096\,$\tablenotemark{a} & $+$1.215 &
	$154 \pm 28$ & $<$ 19100 & $>$ 7.60 \\
NGC 1560--1 & $(1.245 \pm 0.045) \times 10^{-14}$ & \nodata &
	$+$0.354$\,$\tablenotemark{c} & \nodata$\,$\tablenotemark{d} &
	$<$ 11900 & $>$ 8.05 \\
NGC 1560--2 & $(1.725 \pm 0.059) \times 10^{-14}$ & \nodata &
	$+$0.354$\,$\tablenotemark{c} & $111 \pm 11$ &
	$<$ 13300 & $>$ 7.89 \\
NGC 1560--3 & $(7.73 \pm 0.54) \times 10^{-15}$ & \nodata &
	$+$0.354$\,$\tablenotemark{c} & $460 \pm 370$ &
	$<$ 19900 & $>$ 7.55 \\
NGC 1560--4 & $(1.450 \pm 0.077) \times 10^{-14}$ & \nodata &
	$+$0.354$\,$\tablenotemark{c} & \nodata$\,$\tablenotemark{d} &
	$<$ 17300 & $>$ 7.64 \\
NGC 1560--5 & $(8.26 \pm 0.49) \times 10^{-15}$ & \nodata &
	$+$0.354$\,$\tablenotemark{c} & \nodata$\,$\tablenotemark{d} &
	$<$ 18900 & $>$ 7.46 \\
NGC 1560--6 & $(5.87 \pm 0.62) \times 10^{-15}$ & \nodata &
	$+$0.354$\,$\tablenotemark{c} & $8.12 \pm 0.71$ & 
	\nodata & \nodata \\
NGC 1560--7 & $(1.47 \pm 0.29) \times 10^{-15}$ & \nodata &
	$+$0.354$\,$\tablenotemark{c} & $7.7 \pm 1.3$ &
	\nodata & \nodata \\
NGC 3109--1 & $(4.8 \pm 1.2) \times 10^{-16}$ & $+0.15 \pm 0.57$ & 0 &
	\nodata$\,$\tablenotemark{d} & \nodata & \nodata \\
NGC 3109--2 & $(2.47 \pm 0.17) \times 10^{-15}$ & $+0.06 \pm 0.44$ & 0 &  
	$149 \pm 35$ & \nodata & \nodata \\
NGC 3109--3 & $(4.46 \pm 0.20) \times 10^{-15}$ & $-0.03 \pm 0.23$ & 0 &
	$60.2 \pm 2.6$ & $14600 \pm 2600$ & $7.73 \pm 0.33$ \\
NGC 3109--4 & $(1.027 \pm 0.048) \times 10^{-14}$ & $-0.02 \pm 0.29$ & 0 &
	$25.65 \pm 0.80$ & $<$ 16200 & $>$ 7.62 \\
UGC 6456--1 & $(4.99 \pm 0.30) \times 10^{-15}$ & $-0.07 \pm 0.27$ & 0 &
	$66.1 \pm 7.6$ & $20900 \pm 4000$ & $7.21 \pm 0.23$ \\
UGC 6456--2 & $(2.156 \pm 0.086) \times 10^{-14}$ & $-0.18 \pm 0.17$ & 0 &
	$147 \pm 10$ & $17600 \pm 1700$ & $7.45 \pm 0.14$ \\
DDO 187--1 & $(1.937 \pm 0.072) \times 10^{-15}$ & 0 & 0 &
	73$\,$\tablenotemark{e} & $13200 \pm 700$ & $7.69 \pm 0.09$ \\
DDO 187--2 & $(5.88 \pm 0.28) \times 10^{-16}$ & 0 & 0 &
	25$\,$\tablenotemark{e} & \nodata & \nodata \\
\enddata
\tablenotetext{a}{
From \halpha\ and \hbeta\ measurements at OAN;
see Table~\ref{table_ic10_oan}.
}
\tablenotetext{b}{
IC~10--2 (HL111c; Table~\ref{table_ic10_h2rs}): this value of the
oxygen abundance agrees with the value independently derived by
\cite{richer01} for the same Steward data; they obtain 
12$+$log(O/H) = $8.23 \pm 0.09$. 
}
\tablenotetext{c}{
Average $E(B-V)$ computed from \halpha\ and \hbeta\ measurements 
from red spectra; see Table~\ref{table_n1560_reddata}.
}
\tablenotetext{d}{
Negative equivalent width, owing to negative continuum.
}
\tablenotetext{e}{
From \cite{vanzee97}.
}
\tablecomments{
Column (1): \hii\ region.
Column (2): \hbeta\ intensity, corrected for underlying Balmer
absorption and the adopted reddening in Col. (4).
Columns (3) and (4): Computed and adopted reddening values.
Column (5): Observed \hbeta\ emission equivalent width.
Column (6): Computed O$^{+2}$ electron temperature, assuming an electron
density of 100~cm$^{-3}$. 
Column (7): Derived oxygen abundance.
}
\end{deluxetable}


\begin{deluxetable}{ccccccccc}
\renewcommand{\arraystretch}{1.}
\tablecolumns{9}
\tabletypesize{\scriptsize} 
\tablewidth{0pt}
\tablecaption{
Stars and gas in field dIs.
\label{table_alldi_starsgas}}
\tablehead{
\colhead{dI Name} & \colhead{$M_B$} & \colhead{log $M_{\rm H I}$} &
\colhead{log $M_{\rm gas}$} & \colhead{log $M_{\rm H I}/L_B$} &
\colhead{log $M_{\ast}$} & \colhead{$M_{\ast}/L_B$} & \colhead{$\mu$}
& \colhead{log log (1/$\mu$)} \\ 
\colhead{} & \colhead{(mag)} & \colhead{(\msun)} & \colhead{(\msun)} &
\colhead{(\msun/\lsun)} & \colhead{(\msun)} & \colhead{(\msun/\lsun)}
& & \\
\colhead{(1)} & \colhead{(2)} & \colhead{(3)} & \colhead{(4)} &
\colhead{(5)} & \colhead{(6)} & \colhead{(7)} & \colhead{(8)} &
\colhead{(9)}
}
\startdata
    DDO 187 & $-12.72$ & 7.32 & 7.46 & $+$0.044 & 7.03 & 0.56 & 0.729 & $-$0.863 \\ 
       GR 8 & $-12.19$ & 7.04 & 7.17 & $-$0.027 & 6.74 & 0.47 & 0.731 & $-$0.867 \\ 
      Ho II & $-15.98$ & 8.93 & 9.06 & $+$0.344 & 8.55 & 0.91 & 0.767 & $-$0.939 \\ 
      IC 10 & $-15.85$ & 8.14 & 8.27 & $-$0.394 & 8.65 & 1.29 & 0.299 & $-$0.280 \\ 
    IC 1613 & $-14.53$ & 7.97 & 8.10 & $-$0.036 & 8.13 & 1.33 & 0.486 & $-$0.505 \\ 
    IC 2574 & $-17.06$ & 9.16 & 9.30 & $+$0.146 & 8.94 & 0.84 & 0.693 & $-$0.799 \\ 
    IC 4662 & $-15.84$ & 8.40 & 8.53 & $-$0.132 & 8.30 & 0.58 & 0.632 & $-$0.701 \\ 
      Leo A & $-11.53$ & 6.99 & 7.12 & $+$0.186 & 6.45 & 0.44 & 0.813 & $-$1.046 \\ 
        LMC & $-17.94$ & 8.82 & 8.96 & $-$0.544 & 9.37 & 1.01 & 0.279 & $-$0.257 \\ 
   NGC 1560 & $-16.37$ & 8.85 & 8.98 & $+$0.107 & 8.73 & 0.98 & 0.641 & $-$0.714 \\ 
   NGC 1569 & $-16.54$ & 7.99 & 8.13 & $-$0.818 & 8.57 & 0.58 & 0.264 & $-$0.238 \\ 
   NGC 2366 & $-16.28$ & 8.95 & 9.08 & $+$0.243 & 8.78 & 1.18 & 0.670 & $-$0.760 \\ 
   NGC 3109 & $-15.30$ & 8.94 & 9.07 & $+$0.624 & 8.30 & 0.97 & 0.855 & $-$1.168 \\ 
   NGC 4214 & $-18.04$ & 9.24 & 9.37 & $-$0.169 & 9.39 & 0.97 & 0.489 & $-$0.507 \\ 
   NGC 5408 & $-15.81$ & 8.25 & 8.38 & $-$0.269 & 8.49 & 0.94 & 0.439 & $-$0.446 \\ 
     NGC 55 & $-18.28$ & 9.18 & 9.31 & $-$0.326 & 9.52 & 1.04 & 0.383 & $-$0.380 \\ 
   NGC 6822 & $-14.95$ & 8.13 & 8.26 & $-$0.042 & 8.19 & 1.05 & 0.541 & $-$0.574 \\ 
  Sextans A & $-14.04$ & 8.03 & 8.16 & $+$0.219 & 7.64 & 0.69 & 0.767 & $-$0.939 \\ 
  Sextans B & $-14.02$ & 7.65 & 7.78 & $-$0.150 & 7.73 & 0.86 & 0.530 & $-$0.559 \\ 
        SMC & $-16.56$ & 8.95 & 9.09 & $+$0.136 & 8.92 & 1.28 & 0.593 & $-$0.644 \\ 
   UGC 6456 & $-13.90$ & 7.90 & 8.04 & $+$0.154 & 7.58 & 0.68 & 0.741 & $-$0.886 \\ 
        WLM & $-13.92$ & 7.79 & 7.93 & $+$0.033 & 7.65 & 0.78 & 0.652 & $-$0.735 \\
\enddata
\tablecomments{
Column (1): Name of the dI.
Column (2): Absolute magnitude in $B$.
Columns (3) and (4): \hi\ gas and total gas masses.
Column (5): \hi\ gas mass--to--blue--light ratio.
Column (6): Stellar mass.
Column (7): Stellar mass--to--light ratio in $B$.
Column (8): Fraction of mass in baryons in the form of gas.
Column (9): Inverse gas fraction, as conveyed by log~log~(1/$\mu$).
}
\end{deluxetable}


\begin{deluxetable}{cccc}
\renewcommand{\arraystretch}{1.}
\tablecolumns{4}
\tabletypesize{\footnotesize} 
\tablewidth{0pt}
\tablecaption{
Sensitivity to Ingredients in the Two--Component Method.
\label{table_const_mstarlb}}
\tablehead{
\colhead{Model} & \colhead{Note} & \colhead{Obtained Fit} & 
\colhead{RMS in log(O/H) (dex)}
}
\startdata
(1) & $(M_{\ast}/L_B)_{\rm old} = 3.11$ & 
${\cal Y} = (8.55 \pm 0.43) + (1.38 \pm 0.41) {\cal X}$ &
0.250 \\
(2) & $(M_{\ast}/L_B)_{\rm old} = 1.54$ & 
${\cal Y} = (8.71 \pm 0.45) + (1.19 \pm 0.38) {\cal X}$ &
0.248 \\
(3) & $(B-V)_{\rm old} = 0.74$ & 
${\cal Y} = (8.56 \pm 0.40) + (0.92 \pm 0.14) {\cal X}$ &
0.167 \\
(4) & $(M_{\ast}/L_B) = 1$ & 
${\cal Y} = (8.65 \pm 0.44) + (1.16 \pm 0.32) {\cal X}$ &
0.244 \\
(5) & $(M_{\ast}/L_B)_{\rm old} \propto (L_{B,{\rm old}})^{0.175}$ &
${\cal Y} = (8.64 \pm 0.40) + (1.01 \pm 0.17) {\cal X}$ &
0.162 \\
\enddata
\tablecomments{
$\cal Y$ represents the oxygen abundance, 12$+$log(O/H),
and $\cal X$ represents $\log\,\log (1/\mu)$.
Model (1): the mass--to--light ratio for old stars is fixed at a
constant value, based upon the properties of the Milky Way disk.
Model (2): the mass--to--light ratio for old stars is fixed at a
constant mean value over all dIs in the sample.
Model (3): the mass--to--light ratio for old stars varies with the
luminosity of the old component as a power law, and the $B-V$ colour
for old stars is fixed at a constant mean value over all dIs in the
sample.
Model (4): the mass--to-light ratio for {\em all\/} stars is fixed 
at unity, motivated by the \cite{bc96} models
and the absence of any trend in $B-V$ colour with $B$ luminosity
\citep{lee01}.
Model (5): full two--component model (Equation~(\ref{eqn_zmu_fit}). 
}
\end{deluxetable}

\end{document}